\documentclass[12pt,preprint,iop,apj]{emulateapj}  

\usepackage{graphicx}
\usepackage{txfonts}
\newcommand{\va}{v_{\mathrm{A}}}
\newcommand{\vae}{v_{\mathrm{A,e}}}
\newcommand{\vai}{v_{\mathrm{A,i}}}

\newcommand{\der}{{\rm d}}

\newcommand{\rhoi}{\rho_{\rm i}}
\newcommand{\rhoe}{\rho_{\rm e}}
\newcommand{\xii}{\mbox{\boldmath{$\xi$}}}

\newcommand{\rhotr}{\rho_{\rm tr}}

\newcommand{\ra}{r_{\mathrm{A}}}

\usepackage{natbib}

\begin{document}

	\title{The behavior of transverse  waves in  nonuniform solar flux tubes.\\ I. Comparison of ideal and resistive results}

	\shorttitle{Transverse  waves. I. Ideal versus resistive results}

  \author{Roberto Soler$^1$, Marcel Goossens$^2$, Jaume Terradas$^1$, and Ram\'on Oliver$^1$}
     \affil{$^1$Departament de F\'isica, Universitat de les Illes Balears,
               E-07122, Palma de Mallorca, Spain}
  \affil{$^2$Centre for Mathematical Plasma Astrophysics, Department of Mathematics, KU Leuven,
             Celestijnenlaan 200B, 3001 Leuven, Belgium}
              \email{roberto.soler@uib.es}

  \begin{abstract}
Magnetohydrodynamic (MHD) waves are ubiquitously observed in the solar atmosphere. Kink waves are a type of transverse MHD waves in magnetic flux tubes that are  damped due to resonant absorption.  The  theoretical study of kink MHD waves in solar flux tubes is usually based on the simplification that the transverse variation of density is confined to a nonuniform layer much thinner than the radius of the tube, i.e., the so-called thin boundary approximation.  Here, we develop a general analytic method to compute the dispersion relation and the eigenfunctions of ideal MHD waves in pressureless flux tubes with transversely nonuniform layers of arbitrary thickness. Results for kink waves are produced and are compared with fully numerical resistive MHD eigenvalue computations in the limit of small resistivity. We find that the frequency and resonant damping rate are the same in both ideal and resistive cases. The actual results for thick nonuniform layers deviate from the behavior predicted in the thin boundary approximation and strongly depend on the shape of the nonuniform layer. The eigenfunctions  in ideal MHD  are very different from those in resistive MHD. The ideal eigenfunctions display a global character regardless of the thickness of the nonuniform layer, while the resistive eigenfunctions are localized around the resonance and are indistinguishable from those of ordinary resistive Alfv\'en modes. Consequently, the spatial distribution of wave energy in the ideal and resistive cases is dramatically different. This poses a fundamental theoretical problem with clear observational consequences.
  \end{abstract}

   \keywords{Sun: oscillations ---
                Sun: atmosphere ---
		Sun: magnetic fields ---
		waves ---
		Magnetohydrodynamics (MHD)}


\section{INTRODUCTION}

Transverse magnetohydrodynamic (MHD) waves are routinely observed in the solar atmosphere after the first detection of standing waves with TRACE \citep[e.g.,][]{nakariakov1999,aschwanden1999} and of propagating waves with Hinode/SOT and CoMP \citep[e.g.,][]{depontieu2007,tomczyk2007}. Kink waves are a specific type of almost incompressible transverse MHD waves in magnetic flux tubes \citep[see, e.g.,][]{edwin1983,goossens2009,goossens2012}, which are damped by resonant absorption due to naturally occurring plasma inhomogeneity in the direction transverse to the magnetic field \citep[e.g.,][]{goossens1992,goossens2002,rudermanroberts2002,pascoe2010,terradas2010,soler2011strat}.

The theoretical  study of kink MHD waves in solar flux tubes using normal modes usually relies on two simplifications: (1) the thin tube (TT) approximation, that assumes that the wavelength is much longer than the radius of the flux tube, and (2) the thin boundary (TB) approximation, that confines the transverse variation of density to a layer much thinner than the radius of the tube. The combined use of these two approximations is called the TTTB approximation. This method does not make a distinction between standing and propagating waves. In the TTTB approximation,   the period/wavelength is unaffected by the thickness of the nonuniform layer, while the temporal/spatial damping rate is linearly proportional to the thickness of the layer \citep[e,g.,][]{rudermanroberts2002,goossens2002,terradas2010}. A numerical factor  in the  formula for the damping rate is the only remnant of the specific form of the spatial variation of density in the nonuniform layer.  The use of the TT approximation is quite reasonable in view that the magnetic waveguides in the solar atmosphere, e.g., coronal loops, chromospheric spicules, and prominence threads, are usually very thin. For instance, the geometrical properties of oscillating coronal loops reported by \citet{aschwanden2002} indicate that oscillating loops are roughly two orders of magnitude longer than their radii. Conversely, there is no observational justification for the use of the TB approximation.  The only justification for the use of the TB approximation is that it is mathematically convenient to obtain a simple analytic expression for the damping rate. Indeed, some studies suggest that coronal loops are largely inhomogeneous in the transverse direction \citep[see, e.g.,][]{aschwanden2003}, hence the TB approximation may not be realistic.

The study of the normal modes in tubes with thick nonuniform layers requires, in general, the use of numerical computations.  The inclusion of dissipation as, e.g., magnetic resistivity, is needed for the numerical schemes to properly treat the behavior of the wave perturbations across the Alfv\'en resonance. The effect of thick nonuniform layers was first investigated by \citet{vandoorsselaere2004}, who  abandoned the analytic TTTB approximation and obtained the period and damping rate of standing kink waves by means of fully numerical eigenvalue computations in resistive MHD.  Later, \citet{arregui2005} performed a similar study but including longitudinal density stratification. Resistive eigenvalue computations have been subsequently used in various studies \citep[e.g.,][among others]{terradas2006,arregui2008,arregui2011,soler2009slow,soler2009PI}. \citet{vandoorsselaere2004} concluded that the error due to the TTTB approximation when used beyond the limit of thin layers is 25\% at most. Since 25\%  is a relatively small error, the results of \citet{vandoorsselaere2004} support the generalization of the TTTB approximation beyond its theoretical range of applicability. This gave rise to analytic seismological inversion schemes for kink waves that make extensive use of this approximation \citep{goossens2008seis,goossens2012seis} and are much simpler than fully numerical inversions \citep{arregui2007seis}.

In the present paper, we revisit the theoretical investigation of transverse waves in flux tubes with thick nonuniform layers. Our reasons for tackling this task are the following. (1) We aim to find an alternative method to obtain the ideal MHD modes in tubes with thick layers, which does not involve the numerical solution of resistive eigenvalues. This would remove the necessity of using resistive MHD computations. Also, the comparison of the results obtained in ideal MHD with those in resistive MHD would allow us to isolate possible effects introduced by resistivity. (2) We will investigate the kink wave eigenfunctions and energy distribution in tubes with thick layers. \citet{vandoorsselaere2004} focused on studying the period and damping rate and did not investigate the eigenfunctions. The form of the eigenfunctions is important for the spatial distribution of wave energy \citep{goossens2013}.  (3) It is likely that the form of the spatial variation of density plays a role when the nonuniform layer is thick so that most part of the tube is nonuniform. However, seismological inversion schemes based on the TTTB approximation \citep{goossens2008seis,goossens2012seis} use an ad hoc variation of density in the nonuniform layer and ignore the influence of other density profiles.  The effect of the specific transverse density variation on the accuracy of the TTTB approximation has not been determined, so that the impact of the shape of the transitional layer on seismological inversions is unknown.

To achieve the three objectives given above, we  develop a general analytic method to compute the dispersion relation and the eigenfunctions of ideal MHD waves in pressureless flux tubes with transversely nonuniform layers of arbitrary thickness. The analytic process uses the Method of Frobenius to express the solution for the total pressure perturbation in the nonuniform layer as a combination of a singular and a regular series around the Alfv\'en resonance position. The analytic treatment is inspired by the work of \citet{hollweg1990a} on the absorption of MHD waves launched towards a thick Cartesian interface between two plasmas \citep[see also][]{zhu1988,hollweg1990b,wright1994}. The technique allows to consider any variation of density in the nonuniform layer. We investigate kink waves as an application of the method.

In this article, we are concerned with objectives (1) and (2). We present the mathematical method and compare the results in ideal MHD with those in resistive MHD. We investigate the impact of thick transitional layers on the frequency, damping rate, eigenfunctions, and energy distribution of kink waves.  Objective (3) will be addressed in a forthcoming second part of this work, where we will study the effect of the shape of the transitional layer on the accuracy of the TTTB approximation and will explore the implications of the results for seismological inversions.

This article is organized as follows. Section~\ref{sec:model} contains the description of the equilibrium configuration and the basic equations. The mathematical method is presented and explained in Section~\ref{sec:math}. Approximate results for thin nonuniform layers are obtained in Section~\ref{sec:approxthin}. Later, general results for thick nonuniform layers are given in Section~\ref{sec:thick}. The effect of the transverse density profile is investigated in Section~\ref{sec:profile}. Finally, we discuss in Section~\ref{sec:conclusion} the theoretical and observational implications of the results.

\section{MODEL AND BASIC EQUATIONS}
\label{sec:model}

The equilibrium configuration is made of a straight magnetic cylinder of radius $R$ embedded in a uniform and infinite plasma. We use cylindrical coordinates, namely $r$, $\varphi$, and $z$ for the radial, azimuthal, and longitudinal coordinates, respectively. The magnetic field is straight and constant and along the axis of the cylinder, namely ${\bf B} = B {\bf 1}_z$. We adopt the $\beta = 0$ approximation, where $\beta$ refers to the ratio of the gas pressure to the magnetic pressure. This is an appropriate approximation to describe transverse MHD waves in the solar corona. In the $\beta = 0$ approximation we can freely choose the density profile in the equilibrium state. Hence, the density, $\rho$, is chosen uniform in the azimuthal and longitudinal directions and nonuniform in the radial direction, namely $\rho = \rho(r)$. We consider the following profile,
\begin{equation}
 \rho(r) = \left\{
\begin{array}{lll}
\rhoi, & \textrm{if} & r \leq R - l/2, \\
\rhotr(r), & \textrm{if} & R -l/2 < r < R +  l/2,\\
\rhoe, & \textrm{if} & r \geq R+l/2,
\end{array}
\right.
\end{equation}
where $\rhoi$ and $\rhoe$ are internal and external constant densities and $\rhotr(r)$ represents a density profile that continuously connects the internal plasma to the external plasma by a nonuniform transitional layer of arbitrary thickness, $l$. We make no assumption concerning the thickness of the transitional layer. The limit $l/R = 0$ corresponds to an abrupt jump in density, while the case $l/R = 2$ corresponds to a tube fully inhomogeneous in the radial direction. We set $\rhoi > \rhoe$ corresponding to an overdense tube with respect to the external plasma. At the present stage we do not specify the form of $\rhotr(r)$.

The equilibrium magnetic flux tube described above acts as a waveguide for MHD waves. We study linear ideal MHD waves superimposed on the equilibrium state.  We consider no equilibrium flows. We linearize the ideal MHD equations for a pressureless static plasma and the resulting equations are
 \begin{eqnarray}
 \rho(r) \frac{\partial^2 \xii}{\partial t^2} &=& \frac{1}{\mu} \left( \nabla \times {\bf b} \right) \times {\bf B}, \label{eq:mom}\\
{\bf b} &=& \nabla \times \left( \xii \times {\bf B} \right), \label{eq:induc}
\end{eqnarray}
where $\xii = (\xi_r,\xi_\varphi,\xi_z)$ is the plasma Lagrangian displacement, ${\bf b} = (b_r,b_\varphi,b_z)$ is the magnetic field perturbation, and $\mu$ is the magnetic permittivity.

In the present investigation we study normal modes. The temporal dependence of perturbations is put proportional to $\exp(-i \omega t)$, with $\omega$ the frequency. In the stationary state of linear wave propagation any wavepacket can be decomposed as a sum of normal modes with different frequencies. In addition, since the equilibrium  is uniform in both $\varphi$- and $z$-directions, we can restrict ourselves to study the individual Fourier components of the perturbations along these directions. Hence the perturbations are put proportional to $\exp(i m \varphi + i k_z z)$, where $m$ and $k_z$ and the azimuthal and longitudinal wavenumbers, respectively. Only integer values of $m$ are possible. Kink waves are characterized by $m=\pm 1$. From here on, we only retain the dependence of the perturbations on the radial coordinate.

Equations~(\ref{eq:mom}) and (\ref{eq:induc}) can be combined to obtain a differential equation for the total pressure Eulerian perturbation, $P' = {\bf B}\cdot{\bf b}/\mu$, as
\begin{eqnarray}
\frac{\partial^2 P'}{\partial r^2} &+& \left[ \frac{1}{r} - \frac{\frac{\der}{\der r} \left( \rho(r)\left( \omega^2 - k_z^2 \va^2(r) \right) \right)}{\rho(r)\left( \omega^2 - k_z^2 \va^2(r) \right)} \right] \frac{\partial P'}{\partial r} \nonumber \\
&+& \left( \frac{\rho(r)\left( \omega^2 - k_z^2 \va^2(r) \right) }{B^2/\mu} - \frac{m^2}{r^2} \right) P' = 0, \label{eq:ptot}
\end{eqnarray}
where $\va^2(r) = B^2 /\mu \rho(r)$ is the square of the local Alfv\'en velocity. 
The radial and azimuthal components of the Lagrangian displacement, $\xi_r$ and $\xi_\varphi$, are related to $P'$ as
\begin{eqnarray}
\xi_r &=& \frac{1}{\rho(r)\left( \omega^2 - k_z^2 \va^2(r) \right)}\frac{\partial P'}{\partial r}, \label{eq:xir} \\
\xi_\varphi &=& \frac{1}{\rho(r)\left( \omega^2 - k_z^2 \va^2(r) \right)} \frac{im}{r} P'.  \label{eq:xif}
\end{eqnarray}
The longitudinal component of the Lagrangian displacement is $\xi_z =0$  because there are no motions along the magnetic field in the $\beta = 0$ approximation.

Note in Equation~(\ref{eq:ptot}) the presence of the term with the radial derivative of $\rho(r)\left( \omega^2 - k_z^2 \va^2(r) \right)$. This term is zero when the density, $\rho$, is uniform. Conversely, when $\rho$ is nonuniform, this term causes Equation~(\ref{eq:ptot}) to be singular at the specific position in the equilibrium, $r=\ra$, where the resonant condition $\omega^2 = k_z^2 \va^2(\ra)$ is satisfied, with  $\ra$ denoting the Alfv\'en resonance position. This causes the damping of the MHD waves with $m\neq 0$.

We recall that this work is based on normal modes. It is expected that the normal modes determine the behavior of the flux tube oscillations after a transitory phase following the initial excitation \citep[see the numerical simulations by, e.g.,][]{terradas2006rs}. The resonant damping of normal modes follows an exponential law \citep[e.g.,][]{goossens2002,rudermanroberts2002,terradas2010}. Recently, it has been shown in time-dependent numerical simulations that the normal mode exponential damping is preceded, in the first stages of the oscillation, by a Gaussian-like damping phase \citep{pascoe2012,pascoe2013,hood2013,rudermanterradas2013}. The results discussed in this paper apply to the oscillation and damping regimes described by  normal modes.

\section{MATHEMATICAL METHOD}
\label{sec:math}

\subsection{Solution in the uniform regions}

In the regions with constant density  Equation~(\ref{eq:ptot}) becomes
\begin{equation}
\frac{\der^2 P'}{\der r^2} + \frac{1}{r}  \frac{\der P'}{\der r} + \left( \frac{\omega^2 - k_z^2\va^2}{\va^2} - \frac{m^2}{r^2} \right) P' = 0, \label{eq:bessel}
\end{equation}
where now $\va^2$ is constant. Equation~(\ref{eq:bessel}) is the Bessel Equation. We look for solutions to Equation~(\ref{eq:bessel}) in the internal and external plasmas. We use the subscripts `i' and `e' to denote quantities related to the internal and external plasmas, respectively.

In the internal plasma, i.e.,  $r \leq R - l/2$, we require that $P'$ be regular at $r=0$. Thus, 
\begin{equation}
P'_{\rm i} = A_{\rm i} J_{m}\left( k_{\perp,\rm i} r  \right), \label{eq:pin}
\end{equation}
where $A_{\rm i}$ is a constant, $J_m$ is the Bessel function of the first kind of order $m$, and
\begin{equation}
	k_{\perp,\rm i}^2 =  \frac{\omega^2 - k_z^2\vai^2}{\vai^2}.
\end{equation}
We move to the external plasma, i.e., $r \geq R + l/2$. The requirement that the MHD wave is trapped in the flux tube means that $P'$ must vanish at $r\to \infty$. Hence, we discard leaky waves from the present investigation \citep[see, e.g.,][]{cally1986,cally2003}. The solution to Equation~(\ref{eq:bessel}) for $r \geq R + l/2$ is
\begin{equation}
P'_{\rm e} = A_{\rm e} K_{m}\left( k_{\perp,\rm e} r  \right), \label{eq:pex}
\end{equation}
where again $A_{\rm e}$ is a constant, $K_m$ is the modified Bessel function of the first kind of order $m$, and
\begin{equation}
	k_{\perp,\rm e}^2 =  - \frac{\omega^2 - k_z^2\vae^2}{\vae^2}.
\end{equation}

\subsection{Solution in the nonuniform layer}

Here we connect the solution in the internal plasma (Equation~(\ref{eq:pin})) to the solution in the external medium (Equation~(\ref{eq:pex})) by solving Equation~(\ref{eq:ptot}) in the nonuniform transitional layer. We use the subscript `tr' to denote quantities related to the transitional layer. The position of the Alfv\'en resonance, $\ra$, is a regular singular point.  We take advantage of this fact and use the Method of Frobenius to obtain the solution to Equation~(\ref{eq:ptot}) as an infinite power series expansion  around the resonance position \citep[see, e.g.,][]{zhu1988,hollweg1990a,hollweg1990b,wright1994,cally2010}. We assume that there is only one resonance position. The method is outlined in the following paragraphs.

We perform the change of variable
\begin{equation}
\zeta \equiv  r - \ra.
\end{equation}
In this new radial coordinate, the boundaries of the transitional layer are at the positions $\zeta_{\rm i}$ and $\zeta_{\rm e}$, namely
\begin{eqnarray}
\zeta_{\rm i} &=& R - \frac{l}{2} - \ra, \\
\zeta_{\rm e} &=& R + \frac{l}{2} - \ra.
\end{eqnarray}  
We rewrite Equation~(\ref{eq:ptot}) as
\begin{equation}
\zeta^2 h(\zeta) \frac{\partial^2 P'}{\partial \zeta^2} +\zeta p(\zeta)  \frac{\partial P'}{\partial \zeta} + q(\zeta)  P' = 0, \label{eq:ptrfro}
\end{equation}
where the functions $h(\zeta)$, $p(\zeta)$, and $q(\zeta)$ are defined as
\begin{eqnarray}
h(\zeta) &=& \left( \zeta + \ra \right)^2 f(\zeta),\\
p(\zeta) & =&\zeta \left( \zeta + \ra \right)  \left[ f(\zeta)-\left( \zeta + \ra \right) \frac{\partial f(\zeta)}{\partial \zeta} \right], \\
q(\zeta) &=&  \zeta^2 \left[ \frac{\mu}{B^2} \left( \zeta + \ra \right)^2 f^2(\zeta) - m^2 f(\zeta)\right],
\end{eqnarray}
with
\begin{equation}
f(\zeta) = \rho(\zeta) \left( \omega^2 - k_z^2 \va^2(\zeta) \right) = \omega^2\rho(\zeta) - k_z^2 \frac{B^2}{\mu}.
\end{equation}
We assume that the density profile is an analytic function at the resonance position. Hence we perform a Taylor series of $\rho(\zeta)$ around $\zeta=0$ as 
\begin{equation}
\rho(\zeta) = \sum_{k=0}^{\infty} \rho_{k} \zeta^k,
\end{equation}
with $\rho_0 = \rho(\zeta=0) = \rho(r=\ra)$ and
\begin{equation}
\rho_{k} =\frac{1}{k!} \left. \frac{\der^k \rho(\zeta)}{\der \zeta^k} \right|_{\zeta=0} = \frac{1}{k!} \left. \frac{\der^k \rho(r)}{\der r^k} \right|_{r=\ra}, \qquad \textrm{for} \quad k \geq 1.
\end{equation}
Note that we do not specify the form of the density profile, $\rho(r)$. We recall that the only requirements are, first, that the density profile is an analytic function at $r =\ra$ and, second, that there is only one resonance position. The analysis below is valid for any density profile satisfying these two conditions.

We express the solution to Equation~(\ref{eq:ptrfro}) in the form of series expansion around the regular singular point $\zeta=0$, namely
\begin{equation}
P'_{\rm tr}(\zeta) = \zeta^s \sum_{k=0}^{\infty} a_k \zeta^k \label{eq:frob1}
\end{equation}
where $s$ is the index of the expansion and $a_k$ are coefficients to be determined. The value of the coefficients $a_k$ depend on the specific density profile considered. We substitute Equation~(\ref{eq:frob1}) in Equation~(\ref{eq:ptrfro}). The  relation determining the values of the index $s$ is obtained from the coefficient of the lowest power of $\zeta$ in the resulting infinite series. This leads to the indicial equation $s \left( s-2  \right) = 0$, with roots $s_1 = 2$ and $s_2 = 0$ \citep[see also][]{goossens1992}. Then, the general solution to Equation~(\ref{eq:ptrfro}) is the sum of two linearly independent solutions, namely a regular series, $P'_{1}(\zeta)$, and a singular series, $P'_{2}(\zeta)$. The expressions of this two linearly independent solutions are
\begin{eqnarray}
P'_{1}(\zeta) &=& \zeta^2 \sum_{k=0}^{\infty} a_k \zeta^k, \label{eq:regular} \\
P'_{2}(\zeta) &=& \sum_{k=0}^{\infty} s_k \zeta^k + \mathcal{C} P'_{1}(\zeta) \ln \zeta, \label{eq:singular}
\end{eqnarray}
where $\mathcal{C}$ is the coupling constant and $a_k$ and $s_k$ are series coefficients. The general solution to Equation~(\ref{eq:ptrfro}) is then
\begin{equation}
P'_{\rm tr}(\zeta) = A_0 P'_{1}(\zeta) + S_0 P'_{2}(\zeta), \label{eq:seriesgen}
\end{equation}
where $A_0$ and $S_0$ are constants. Since the general solution to a 2nd order ordinary differential equation contains two undetermined coefficients, we adopt $a_0 = s_0 = 1$ with no loss of generality. The expressions of the  remaining coefficients $a_k$ and $s_k$ and that of the coupling constant $\mathcal{C}$ are obtained after substituting Equation~(\ref{eq:seriesgen}) in  Equation~(\ref{eq:ptrfro}). The coefficients $a_k$ and $s_k$  depend on the choice of the density profile, but the coupling constant is independent of the density profile, namely 
\begin{equation}
\mathcal{C} =\frac{m^2}{2\ra^2}.
\end{equation}
When $m=0$, $\mathcal{C} = 0$ and the singular series, $P'_{2}(\zeta)$, becomes a regular series due to the absence of the logarithmic term. As a consequence, no resonance takes place when $m=0$. General expressions of the coefficients $a_k$ and $s_k$ and their recurrence relations are given in the Appendix.

Finally, the expressions of $\xi_r$ and $\xi_\varphi$ in the transitional layer are straightforwardly obtained by substituting Equation~(\ref{eq:seriesgen}) in Equations~(\ref{eq:xir}) and (\ref{eq:xif}), respectively.

\subsection{Dispersion relation}  
  
The  dispersion relation is obtained by imposing  the continuity of $P'$ and $\xi_r$ at $r=R - l/2$ and $r=R + l/2$.  These boundary conditions provide us with a system of four algebraic equations for the constants $A_{\rm i}$, $A_{\rm e}$, $A_0$, and $S_0$. The requirement that there is a nontrivial solution of the system provides us with the dispersion relation.  For simplicity, we omit the intermediate steps and give the final expression of  the dispersion relation, namely
\begin{eqnarray}
&& \frac{\frac{k_{\perp,\rm e}}{\rhoe \left( \omega^2 - k_z^2\vae^2 \right)} \frac{K'_{m}\left[ k_{\perp,\rm e}(R+l/2)\right]}{K_{m}\left[ k_{\perp,\rm e}(R+l/2)\right]}\mathcal{G}_{\rm e}-  \Xi_{\rm e}}{\frac{k_{\perp,\rm e}}{\rhoe \left( \omega^2 - k_z^2\vae^2 \right)} \frac{K'_{m}\left[ k_{\perp,\rm e}(R+l/2)\right]}{K_{m}\left[ k_{\perp,\rm e}(R+l/2)\right]}\mathcal{F}_{\rm e}-  \Gamma_{\rm e}} \nonumber \\
&-&\frac{\frac{k_{\perp,\rm i}}{\rhoi \left( \omega^2 - k_z^2\vai^2 \right)} \frac{J'_{m}\left[ k_{\perp,\rm i}(R-l/2)\right]}{J_{m}\left[ k_{\perp,\rm i}(R-l/2)\right]}\mathcal{G}_{\rm i}- \Xi_{\rm i}}{\frac{k_{\perp,\rm i}}{\rhoi \left( \omega^2 - k_z^2\vai^2 \right)} \frac{J'_{m}\left[ k_{\perp,\rm i}(R-l/2)\right]}{J_{m}\left[ k_{\perp,\rm i}(R-l/2)\right]}\mathcal{F}_{\rm i}-  \Gamma_{\rm i}} = 0, \label{eq:reldisper}
\end{eqnarray}
where
\begin{eqnarray}
\mathcal{G}_{\rm i,e} & = & \sum_{k=0}^\infty a_k \zeta_{\rm i,e}^{k+2}, \\
\mathcal{F}_{\rm i,e} & = & \sum_{k=0}^\infty \left( s_k \zeta_{\rm i,e}^{k} + \frac{m^2}{2\ra^2}\ln\left( \zeta_{\rm i,e} \right) a_k \zeta_{\rm i,e}^{k+2} \right), \\
\Xi_{\rm i,e}  & = &  \frac{1}{\omega^2\sum_{k=0}^\infty \rho_{k+1}\zeta_{\rm i,e}^k}\sum_{k=0}^\infty (k+2)a_k \zeta_{\rm i,e}^{k}, \\\
\Gamma_{\rm i,e} & = & \frac{1}{\omega^2\sum_{k=0}^\infty \rho_{k+1}\zeta_{\rm i,e}^k}\sum_{k=0}^\infty\left(  k s_k \zeta_{\rm i,e}^{k-2} + \frac{m^2}{2\ra^2} a_k \zeta_{\rm i,e}^{k} \right. \nonumber \\
 &&+ \left. \frac{m^2}{2\ra^2} \ln\left( \zeta_{\rm i,e}\right)  (k+2)a_k \zeta_{\rm i,e}^{k}\right).
\end{eqnarray} 

Equation~(\ref{eq:reldisper}) is valid for any value of $m$, namely $m=0$ sausage modes, $m=1$ kink modes, and $m \geq 2$ fluting modes. When $m \neq 0$, the dispersion relation is a multivalued function  due to the presence of logarithmic terms.  To make the dispersion relation univalued when $m\neq 0$, the branch points of the logarithm functions are connected in the complex plane with appropriate branch cuts \citep[see details in, e.g.,][]{tataronis1973a,goedbloed2004}. The dispersion relation has no solutions on the principal Riemann sheet because, strictly speaking, complex eigenvalues are not possible in ideal MHD \citep{poedts1991,goedbloed2004}. To find the  physical solutions that represent damped waves, we have to consider the analytic continuation of the dispersion relation to the next Riemann sheet \citep[see, e.g.,][]{sedlacek1971}. The logarithmic terms are absent when $m=0$, so that the dispersion relation is univalued and there is no resonant damping in that case.

We recall that neither the form of the density variation nor the thickness of the non-uniform layer have been imposed so far. Equation~(\ref{eq:reldisper}) is the valid dispersion relation for any density profile and for $l/R \in (0,2)$. The form of the density profile only affects the values of the coefficients in $\mathcal{G}_{\rm i,e}$, $\mathcal{F}_{\rm i,e}$,   $\Xi_{\rm i,e}$, and $\Gamma_{\rm i,e}$.

Also, note that the present formalism is the same for both standing and propagating waves. The dispersion relation (Equation~(\ref{eq:reldisper})) is the same in both cases. Standing waves are described by a real $k_z$, so that the solution of Equation~(\ref{eq:reldisper}) is a complex frequency, namely $\omega = \omega_{\rm R} + i \omega_{\rm I}$, where $\omega_{\rm R}$ and $\omega_{\rm I}$ are the real and imaginary parts of $\omega$, respectively. Due to resonant damping $\omega_{\rm I} < 0$ and, because of the temporal dependence $\exp(-i\omega t)$, the amplitude of the waves decreases in time by the exponential factor $\exp(-\left|\omega_{\rm I}\right| t)$. Conversely propagating waves are described by a real $\omega$, so that the solution of the dispersion relation is a complex longitudinal wavenumber, namely $k_z = k_{z,\rm R} + i k_{z,\rm I}$, where $k_{z,\rm R}$ and $k_{z,\rm I}$ are the real and imaginary parts of $k_z$, respectively, and $k_{z,\rm I} > 0$. The amplitude of the propagating waves decreases in $z$ by the exponential factor $\exp(-k_{z,\rm I} z)$. Therefore, standing and propagating cases are equivalent and are both described by the same dispersion relation (Equation~(\ref{eq:reldisper})).

\section{APPROXIMATE RESULTS FOR THIN LAYERS}

\label{sec:approxthin}

In general, for arbitrary values of $l/R$, Equation~(\ref{eq:reldisper}) has to be solved numerically. However,  analytic approximations to the solutions with $m\neq 0$ can be obtained when $l/R$ is a small parameter and only the leading terms in the expressions of $\mathcal{G}_{\rm i,e}$, $\mathcal{F}_{\rm i,e}$,   $\Xi_{\rm i,e}$, and $\Gamma_{\rm i,e}$ are retained.  We  assume that the nonuniform layer is sufficiently thin so that it suffices to keep  terms up to $\mathcal{O}(l/R)$. A similar situation has been previously studied by, e.g., \citet{goossens1992,goossens2002,rudermanroberts2002} for standing waves and by \citet{terradas2010} for propagating waves using the TB approximation, i.e., $l/R\ll 1$. Before tackling the general study for arbitrary $l/R$, the purpose of this Section is to recover with the present formalism the known results in the TB limit.  This is useful to check the validity of the Frobenius method.

\subsection{Solution to the dispersion relation}

By  retaining terms up to $\mathcal{O}(l/R)$ only, the expressions of  $\mathcal{G}_{\rm i,e}$, $\mathcal{F}_{\rm i,e}$,   $\Xi_{\rm i,e}$, and $\Gamma_{\rm i,e}$ reduce to
\begin{eqnarray}
\mathcal{G}_{\rm i,e} & \approx & 0, \\
\mathcal{F}_{\rm i,e} & \approx & 1, \\
\Xi_{\rm i,e}  & \approx &  \frac{2}{\omega^2\rho_1} = \frac{2}{\omega^2 \left( \der \rho / \der r \right)_R}, \\
\Gamma_{\rm i,e} & \approx & \frac{m^2/R^2}{\omega^2\rho_1}\left(\ln \zeta_{\rm i,e} + \frac{1}{2} \right) = \frac{m^2/R^2}{\omega^2\left( \der \rho / \der r \right)_R}\left(\ln \zeta_{\rm i,e} + \frac{1}{2} \right) ,
\end{eqnarray} 
where  we  took $\ra\approx R$ as consistent with the assumption that the layer is thin so that the resonance position is close to $r=R$. We approximate $\mathcal{G}_{\rm i,e} \approx 0$  because the first nonzero terms in $\mathcal{G}_{\rm i,e}$ are of $\mathcal{O}(l/R)^2$. Then, the dispersion relation (Equation~(\ref{eq:reldisper})) becomes
\begin{eqnarray}
&&\frac{k_{\perp,\rm e}}{\rhoe\left( \omega^2 - k_z^2 \vae^2 \right)} \frac{K'_{m}\left( k_{\perp,\rm e}R\right)}{K_{m}\left( k_{\perp,\rm e}R\right)} \nonumber \\
&-&  \frac{k_{\perp,\rm i}}{\rhoi\left( \omega^2 - k_z^2 \vai^2 \right)} \frac{J'_{m}\left( k_{\perp,\rm i}R\right)}{J_{m}\left( k_{\perp,\rm i}R\right)} =  \frac{m^2/R^2}{\omega^2\left( \der \rho / \der r \right)_R} \ln\left( \frac{\zeta_{\rm e}}{\zeta_{\rm i}} \right), \label{eq:reldisperTB0}
\end{eqnarray}
where  we used $R-l/2 \approx R+l/2 \approx R$  as consistent with the thin layer assumption. Equation~(\ref{eq:reldisperTB0}) is valid for arbitrary values of $k_zR$ because no restriction has been imposed on the radius of the magnetic tube.  When $l=0$, $\left( \der \rho / \der r \right)_R \to \infty$ and the right-hand side of Equation~(\ref{eq:reldisperTB0}) vanishes. Then, the dispersion relation simplies to
\begin{eqnarray}
&&\frac{k_{\perp,\rm e}}{\rhoe\left( \omega^2 - k_z^2 \vae^2 \right)} \frac{K'_{m}\left( k_{\perp,\rm e}R\right)}{K_{m}\left( k_{\perp,\rm e}R\right)} \nonumber \\
&-&  \frac{k_{\perp,\rm i}}{\rhoi\left( \omega^2 - k_z^2 \vai^2 \right)} \frac{J'_{m}\left( k_{\perp,\rm i}R\right)}{J_{m}\left( k_{\perp,\rm i}R\right)} =0, \label{eq:reldisperl0}
\end{eqnarray}
Equation~(\ref{eq:reldisperl0}) is the well-known dispersion relation of \citet{edwin1983}. To obtain an analytic approximation for kink waves we consider the TT approximation and take the limit $k_zR \ll 1$. We use an asymptotic expansion for small arguments and $m\neq0$ of the Bessel functions in Equation~(\ref{eq:reldisperl0}) and keep the first term in the expansions, so that we approximate
\begin{equation}
 \frac{J'_{m}\left( k_{\perp,\rm i}R\right)}{J_{m}\left( k_{\perp,\rm i}R\right)} \approx \frac{m}{k_{\perp,\rm i}R}, \qquad  \frac{K'_{m}\left( k_{\perp,\rm e}R\right)}{K_{m}\left( k_{\perp,\rm e}R\right)} \approx - \frac{m}{k_{\perp,\rm e}R}. 
\end{equation}
Equation~(\ref{eq:reldisperl0}) simplifies to
\begin{equation}
 \rhoi \left( \omega^2 - k_z^2\vai^2 \right)  +  \rhoe \left( \omega^2 - k_z^2\vae^2 \right) =0.  \label{eq:reldisperTT}
\end{equation}
For standing waves, $k_z$ is fixed and $\omega$ is given by the solution of Equation~(\ref{eq:reldisperTT}), namely
\begin{equation}
 \omega^2 = \frac{\rhoi\vai^2 + \rhoe\vae^2}{\rhoi+\rhoe} k_z^2 = \frac{2 B^2/\mu}{\rhoi + \rhoe } k_z^2 \equiv \omega_{k}^2. \label{eq:wk}
\end{equation}
$\omega_k$ is real and is called the kink frequency \citep[see, e.g.,][]{edwin1983,goossens2009}. Conversely, for propagating waves $\omega$ is fixed and $k_z$ is given by the solution of Equation~(\ref{eq:reldisperTT}), namely
\begin{equation}
 k_z^2 = \frac{\rhoi+\rhoe}{\rhoi\vai^2 + \rhoe\vae^2} \omega^2 = \frac{\rhoi + \rhoe }{2 B^2/\mu} \omega^2 \equiv k_{z,\rm k}^2, \label{eq:kzk}
\end{equation}
where $k_{z,k}$ can be equivalently called the kink wavenumber.

We go back to the general case with $l \neq 0$. To evaluate the logarithmic term on the right-hand side of Equation~(\ref{eq:reldisperTB0}), we realize that assuming $\ra \approx R$ results in $\zeta_{\rm i} \approx -l/2$ and $\zeta_{\rm e} \approx l/2$, and we define the complex logarithm so that it jumps $\pm i\pi$ when crossing the negative real axis. Accordingly we approximate $\ln\left( \zeta_{\rm e} / \zeta_{\rm i} \right) \approx \pm i \pi$, where either the $+$ sign or the $-$ sign is conveniently chosen depending on the sign of $\left( \partial \rho / \partial r \right)_R$. This choice is based on the physical argument that the effect of the resonance is to produce the damping of the waves. As in the case with $l=0$, in order to make further analytic progress we consider the TT approximation, $k_zR \ll 1$, and expand the Bessel functions for small arguments. Equation~(\ref{eq:reldisperTB0}) reduces to
\begin{eqnarray}
 & &\rhoi \left( \omega^2 - k_z^2\vai^2 \right)  +  \rhoe \left( \omega^2 - k_z^2\vae^2 \right)   \nonumber \\
&-& i\pi \frac{m}{R} \frac{\rhoi\rhoe}{\left| \der \rho / \der r \right|_R} \frac{\left( \omega^2 - k_z^2\vai^2 \right)\left( \omega^2 - k_z^2\vae^2 \right)}{\omega^2} = 0. \label{eq:reldisperTB}
\end{eqnarray}
Equation~(\ref{eq:reldisperTB}) agrees with the dispersion relation derived by \citet{goossens1992} in the TT and TB limits. The joint use of the TT and TB approximations is called here the TTTB approximation.

Let us first study standing waves. To find an analytic approximation to the kink mode frequency, we write  $\omega = \omega_{\rm R} + i \omega_{\rm I}$ and fix $k_z$ to a real value. We assume weak damping, i.e., $\omega_{\rm I}^2 \ll \omega_{\rm R}^2$, so that we neglect terms with $\omega_{\rm I}^2$ and higher powers in Equation~(\ref{eq:reldisperTB}). The real part of the frequency, $\omega_{\rm R}$, is approximately obtained by setting the real part of Equation~(\ref{eq:reldisperTB}) to zero and substituting $\omega^2$ by $\omega_{\rm R}^2$. Hence we obtain
\begin{equation}
 \omega_{\rm R} \approx \omega_{k}, \label{eq:wrl}
\end{equation}
where $\omega_k$ is given in Equation~(\ref{eq:wk}). The wave period, $P$, is
\begin{equation}
P = \frac{2 \pi}{\omega_{\rm k}}. \label{eq:p}
\end{equation}
According to Equations~(\ref{eq:wrl}) and (\ref{eq:p}), the kink wave frequency, and so the period, is unaffected by the thickness of the transitional layer in the first-order approximation. The same expression as for  $l= 0$ is found. 

The approximation for $\omega_{\rm I}$ is obtained using the expression
\begin{equation}
\omega_{\rm I} \approx - \left. \frac{D_{\rm I}}{\partial D_{\rm R} / \partial \omega}\right|_{\omega \approx \omega_{\rm R}}, 
\end{equation} 
 where $ D_{\rm R} $ and $ D_{\rm I} $ are the real and imaginary parts of the dispersion relation (Equation~(\ref{eq:reldisperTB})), respectively. After some algebraic manipulations the result is
\begin{equation}
 \omega_{\rm I} \approx - \frac{\pi}{8} \frac{m}{R} \frac{\left(\rhoi - \rhoe\right)^2}{\rhoi + \rhoe} \frac{\omega_k}{\left| \der \rho / \der r \right|_R}. \label{eq:wi}
\end{equation}
Now we express
\begin{equation}
\left| \frac{\der \rho}{ \der r} \right|_R = F \frac{\pi^2}{4} \frac{\rhoi - \rhoe}{l}, \label{eq:densider}
\end{equation}
where we introduce the numerical factor $F$ that depends on the specific form of the density profile. For example, $F = 4/\pi^2$ for a linear variation of density and $F = 2/\pi$ for a sinusoidal variation of density.  Using Equation~(\ref{eq:densider}) in Equation~(\ref{eq:wi}) we obtain
\begin{equation}
 \omega_{\rm I} \approx - \frac{m}{2\pi F} \frac{l}{R} \frac{\rhoi - \rhoe}{\rhoi + \rhoe} \omega_k. \label{eq:wi2}
\end{equation}
 Equation~(\ref{eq:wi2}) is  the same damping rate found previously by, e.g., \citet{goossens1992,goossens2002} for a linear density profile and by \citet{rudermanroberts2002} for a sinusoidal density profile.  Equation~(\ref{eq:wi2}) is also equivalent to the damping rate found by, e.g., \citet{sedlacek1971,ionson1978,lee1986,hollwegyang1988} for a surface wave in a Cartesian interface with a linear density profile, where $m/R$ has to be  replaced by the component of the wavevector in the direction perpendicular to both the interface and the density gradient. 
 
  We compute the damping time as $\tau_{\rm D}  = 1/|\omega_{\rm I}|$ and use  Equations~(\ref{eq:p}) and (\ref{eq:wi2}) to give the expression of the ratio of the damping time to the period, namely
 \begin{equation}
 \frac{\tau_{\rm D}}{P} = \frac{F}{m} \frac{R}{l} \frac{\rhoi+\rhoe}{\rhoi-\rhoe}. \label{eq:tttb}
 \end{equation}

 In summary, Equation~(\ref{eq:wrl}) shows that the kink wave frequency is independent of $l/R$, whereas Equation~(\ref{eq:wi2}) predicts a linear dependence of the damping rate with $l/R$, so that  the larger $l/R$, the stronger the damping. The factor $F$ is the only remnant of the form of the density profile that remains in the TTTB formula. Apart from this numerical factor, the dependence of $\tau_{\rm D}/P$ with $l/R$ is not affected by the form of the density profile in the thin nonuniform layer limit (Equation~(\ref{eq:tttb})).

Now, we turn to propagating waves.  We write  $k_z = k_{z,\rm R} + i k_{z,\rm I}$ and fix $\omega$ to a real value. Following the same process as before, we find the approximations to  $k_{z,\rm R}$ and $k_{z,\rm I}$ as
\begin{eqnarray}
k_{z,\rm R} & \approx & k_{z,k}, \\
k_{z,\rm I} & \approx & \frac{m}{2\pi F} \frac{l}{R} \frac{\rhoi - \rhoe}{\rhoi + \rhoe} k_{z,k},
\end{eqnarray}
which agree with the expressions found by \citet{terradas2010}. We compute the ratio of the damping length, $L_{\rm D}  = 1/k_{z,\rm I}$, to the wavelength, $\lambda=2\pi/k_{z,\rm R}$, as
 \begin{equation}
 \frac{L_{\rm D}}{\lambda} = \frac{F}{m} \frac{R}{l} \frac{\rhoi+\rhoe}{\rhoi-\rhoe}, \label{eq:tttb2}
 \end{equation} 
 which is exactly the same expression as for the standing waves $\tau_{\rm D}/P$ (Equation~(\ref{eq:tttb})). In both standing and propagating cases the Frobenius method is consistent with the approximations found in previous works when the TT ($k_zR\ll 1$) and TB ($l/R\ll 1$) limits are taken in the general dispersion relation.

\subsection{Eigenfunctions and radial energy flux}  
\label{sec:pert2}

Approximations to the eigenfunctions of $P'$, $\xi_r$, and $\xi_\varphi$  in the case of an abrupt jump in density were obtained in the TT limit by, e.g., \citet{dymova2006} \citet{goossens2009}.  In the regions with constant density, we follow these previous works and perform asymptotic expansions for small arguments and $m\neq 0$ of Equations~(\ref{eq:pin}) and (\ref{eq:pex}). In the thin nonuniform layer, we only keep terms up to $\mathcal{O}\left( l/R \right)$ in the general expression of $P'_{\rm tr}$ (Equation~(\ref{eq:seriesgen})). Hence, the approximation to $P'$  is 
\begin{equation}
 P'(r) \approx \left\{
\begin{array}{lll}
S_0 \frac{r}{R}, & \textrm{if} & r \leq R - l/2, \\
S_0, & \textrm{if} &  R - l/2 < r <  R + l/2, \\
S_0 \frac{R}{r}, & \textrm{if} & r \geq R + l/2,
\end{array}
\right.
\end{equation}
The value of $P'$ in the thin transitional layer is constant \citep{hollwegyang1988}. In turn, the approximations for the Lagrangian displacements $\xi_{r}$ and $\xi_{\varphi}$ up to $\mathcal{O}\left( l/R \right)$ are
\begin{equation}
 \xi_r(r) \approx \left\{
\begin{array}{lll}
D, & \textrm{if} & r \leq R - l/2, \\
D +  \frac{m^2}{R^2}  \frac{S_0}{\omega^2\left( \der \rho / \der r \right)_R} \ln\frac{r-\ra}{R-\ra - l/2}  , & \textrm{if} &  R - l/2 < r <  R + l/2, \\
 D \left( \frac{R}{r} \right)^2, & \textrm{if} & r \geq R + l/2,
\end{array}
\right. 
\end{equation}
and
\begin{equation}
 \xi_\varphi(r) \approx \left\{
\begin{array}{lll}
i m D, & \textrm{if} & r \leq R - l/2, \\
\frac{i m}{R} \frac{S_0}{\omega^2\left( \der \rho / \der r \right)_R} \frac{1}{r-\ra}, & \textrm{if} &  R - l/2 < r <  R + l/2, \\
 -i m D \left( \frac{R}{r} \right)^2, & \textrm{if} & r \geq R + l/2,
\end{array}
\right. 
\end{equation}
 where
\begin{equation}
D = \frac{1}{\rhoi\left( \omega^2 - k_z^2 \vai^2 \right)} \frac{S_0}{R} = - \frac{1}{\rhoe\left( \omega^2 - k_z^2 \vae^2 \right)} \frac{S_0}{R}. \label{eq:d}
\end{equation} 
 We obtain that $\xi_{r}$ has a logarithmic jump at $r=\ra$, while $\xi_{\varphi}$ behaves as $\xi_{\varphi}\sim \left( r - \ra \right)^{-1}$ in the transitional layer. This implies that the dominant dynamics in the vicinity of the resonance is contained in the azimuthal component or, in a general geometry, the component in the magnetic surfaces perpendicular to the magnetic field lines \citep[see, e.g.,][]{goedbloed1983}. 

We define the variation of the quantity $X$ across the transitional layer as
\begin{equation}
\left[ X  \right] = X_{\rm e} - X_{\rm i},
\end{equation}
where $X_{\rm e,i}$ denote the value of $X$ at $r = R \pm l/2$, respectively. Thus the variation of $P'$ and $\xi_r$ across the thin transitional layer is
\begin{eqnarray}
\left[ P' \right] &=& P'_{\rm tr}(R + l/2) - P'_{\rm tr}(R-l/2) \approx 0, \label{eq:jumpp} \\
\left[ \xi_r \right] &=& \xi_{r,\rm tr}(R+l/2) - \xi_{r,\rm tr}(R-l/2) \approx \frac{m^2}{R^2} \frac{S_0}{\omega^2\left( \der \rho / \der r \right)_R} \ln \frac{\zeta_{\rm e}}{\zeta_{\rm i}} \nonumber \\
& \approx & - i \pi \frac{m^2/R^2}{\omega^2\left| \der \rho / \der r \right|_R} P'(R). \label{eq:jumpxir}
\end{eqnarray}
The variation of $P'$ and $\xi_r$ across a thin nonuniform layer obtained with the present method coincide with the relations derived by, e.g., \citet{sakurai1991}, for the jump of the eigenfunctions across the resonant layer in the presence of dissipation. These jump relations appear naturally in ideal MHD with the Frobenius method. In the specific case with $l=0$, both $P'$ and $\xi_r$ are continuous at $r=R$, while only $\xi_\varphi$  jumps at the boundary so that $\xi_{\varphi,\rm i} = - \xi_{\varphi,\rm e}$ at $r=R$.

The finite jumps of the eigenfunctions in the nonuniform layer are mathematically caused by the logarithmic term in the singular Frobenius series. The physical reason for the existence of these jumps is that there is a net flux of energy toward the nonuniform layer from both sides,  which is the ultimate cause of the kink wave damping. The expression of the radial component of the time-averaged energy flux is \citep[e.g.,][]{bray1974,stenuit1999,arregui2011,goossens2013} 
\begin{equation}
\left< S_r \right> = - \frac{1}{2} {\rm Re} \left( i \omega \xi_r P'^{*} \right), \label{eq:sr}
\end{equation}
where the asterisk denotes the complex conjugate. We use Equations~(\ref{eq:jumpp}) and (\ref{eq:jumpxir}) to compute the jump of $\left< S_r \right>$ as
\begin{equation}
\left[ \left< S_r \right> \right] \approx - \frac{\pi}{2} \frac{m^2/R^2}{\omega_{\rm R}\left| \der \rho / \der r \right|_R} P'^2(R). \label{eq:jumpsr}
\end{equation}
This expression is in agreement with \citet{andries2000}. Since $\left[ \left< S_r \right> \right] < 0$, the energy inflow into the resonance from the flux tube interior is larger than the energy inflow from the external plasma.  When $l=0$, $\left< S_r \right> = 0$ so that there is no radial flux of energy.

\section{RESULTS FOR THICK LAYERS}
\label{sec:thick}  
  
\begin{figure*}
\centering
\includegraphics[width=.99\columnwidth]{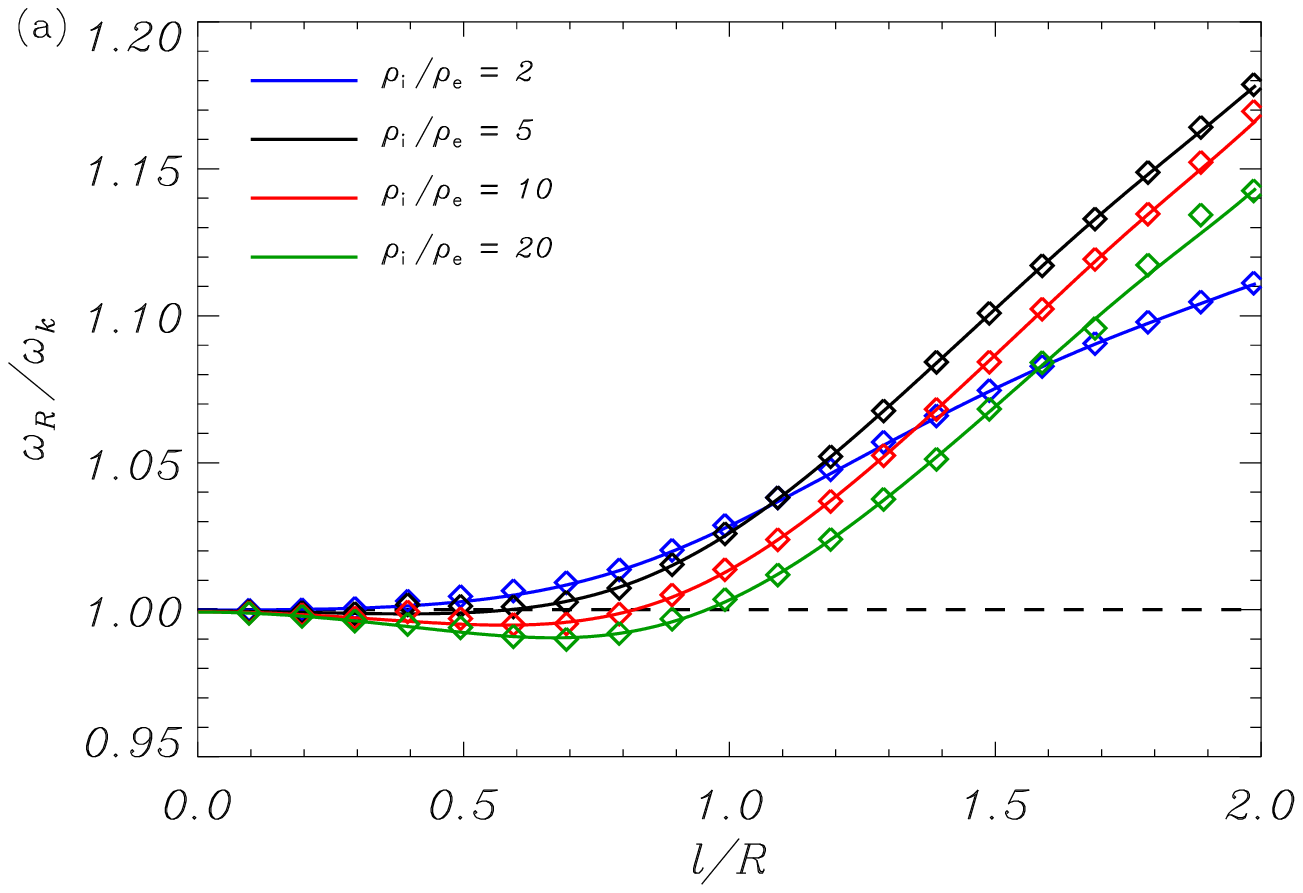}
\includegraphics[width=.99\columnwidth]{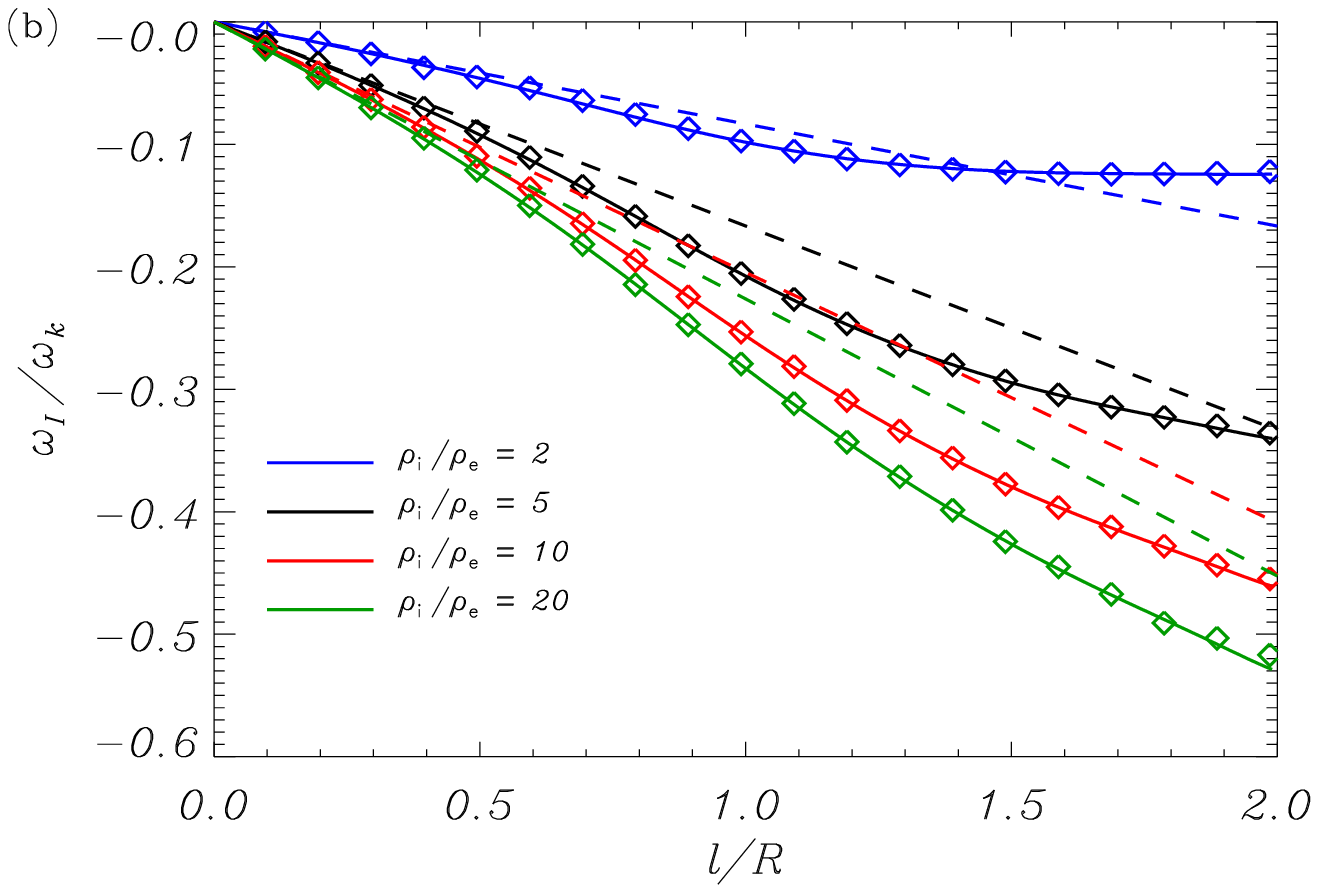}
	\caption{(a) Real part and (b) imaginary part of the kink mode frequency versus $l/R$. In both panels the solid lines correspond to the solutions of the ideal dispersion relation (Equation~(\ref{eq:reldisper})), the dashed lines are the TTTB analytic results (Equations~(\ref{eq:wrl}) and (\ref{eq:wi2})), and the symbols are the resistive MHD eigenvalue results obtained with the PDE2D code. The line color denotes the value of $\rhoi/\rhoe$ used in the computation (indicated within the figures). In all cases we use a sinusoidal transition of density and $L/R = 100$.}
	\label{fig:sin1}
\end{figure*}

In Section~\ref{sec:approxthin} we showed that the solutions obtained with the Frobenius method  consistently revert to the approximations found in previous works when the TTTB limit is considered. Here, we go beyond the limitation of the TTTB approximation. We fully exploit the Frobenius method by considering arbitrary values of the ratio $l/R$. We solve the general dispersion relation (Equation~(\ref{eq:reldisper}))  to compute the frequency and damping rates beyond the case of thin layers. The dispersion relation is a transcendental equation whose roots are found by standard numerical methods. The expressions of $\mathcal{G}_{\rm i,e}$, $\mathcal{F}_{\rm i,e}$,   $\Xi_{\rm i,e}$, and $\Gamma_{\rm i,e}$ involve series with infinite number of terms. To proceed numerically we must truncate the infinite series so that only the first $N$ terms are accounted for. To make sure that the number of terms considered is large enough for the error to be negligible, we  perform convergence tests by increasing $N$ until a good convergence of the solution to Equation~(\ref{eq:reldisper}) is obtained. Typically, we take $N=51$. 

In addition to the ideal normal modes obtained with the Frobenius method, we consider the numerical solution of the resistive eigenvalue problem with the PDE2D code \citep{sewell}. This is the same approach used by \citet{vandoorsselaere2004}, although here we perform the computations with a different numerical code. This will allow us to compare the results in ideal MHD with those in resistive MHD. In the numerical code, the magnetic diffusion term,  $\eta \nabla^2 {\bf b}$, is included in the right-hand side of the linearized induction equation (Equation~(\ref{eq:induc})), where $\eta$ is the coefficient of magnetic resistivity. We take $\eta$ as a constant for simplicity. We define the magnetic Reynolds number as $R_m = \vai R/\eta$. In the solar corona $R_m \sim 10^{12}$ because $\eta$ is extremely small. Using a realistic value for $R_m$ requires taking an enormous number of grid points in the numerical domain. This is not practical from the computational point of view. Therefore, we use in our computations a smaller value of $R_m$. We take values of $R_m$ in between $10^5$--$10^7$. This is computationally advantageous because we can take a smaller number of grid points. However, although in our computations $R_m$ is smaller than its realistic value, we make sure that $R_m$ is still large enough for resistive damping to be negligible compared to resonant damping. In other words, resistivity has no impact on the wave frequency and damping rate \citep[see, e.g.,][]{poedts1991,vandoorsselaere2004,terradas2006}. The PDE2D code solves the resistive eigenvalue problem using a finite-element scheme in a nonuniform grid. The numerical integration of the resistive MHD equations is performed from the cylinder axis, $r = 0$, to the edge of the numerical domain, $r = r_{\rm max}$. We take $r_{\rm max} \gg R$ to avoid numerical errors and to obtain a good convergence of the solution. The PDE2D code uses a collocation method and the generalized matrix eigenvalue problem is solved using the shifted inverse power method. The output of the program is the closest complex eigenvalue to the initial provided guess and the corresponding eigenfunctions.

Since standing and propagating waves are equivalent from the mathematical point of view, from here on we focus on standing waves. We introduce the parameter $L$ that represents the length of the magnetic flux tube. Standing waves whose perturbations are line-tied at the ends of the flux tube are characterized by quantized values of $k_z$ given by
\begin{equation}
k_z = \frac{n\pi}{L},
\end{equation}   
where $n=1$ for the fundamental mode, $n=2$ for the first overtone, and so on. We restrict ourselves to the fundamental kink mode, so we take $n=1$ and $m=1$. In this Section, we assume a sinusoidal variation for the density in the nonuniform layer, namely
\begin{equation}
\rho_{\rm tr}(r) = \frac{\rhoi}{2} \left[\left( 1 + \frac{\rhoe}{\rhoi} \right) -  \left( 1 - \frac{\rhoe}{\rhoi} \right) \sin \left(\frac{\pi}{l}(r-R) \right) \right]. \label{eq:sin}
\end{equation}
This is the same density profile used in the computations by \citet{vandoorsselaere2004}. On purpose, we choose this density profile to compare our results with those reported in this previous work.  Other equilibrium density profiles are considered in Section~\ref{sec:profile}.

\subsection{Frequency and damping rate}
\label{sec:kink1}

Figure~\ref{fig:sin1} displays the real and imaginary parts of the fundamental kink mode frequency versus $l/R$ for a tube with $L/R=100$ and four different values of the density contrast, $\rhoi/\rhoe$. The solid lines correspond to the results of the ideal Frobenius method.

Regarding the real part of the frequency (Figure~\ref{fig:sin1}(a)), we find that $\omega_{\rm R} \approx \omega_k$ is a reasonably good approximation up to $l/R \approx 1$. For thicker layers, the real part of the kink wave frequency increases and becomes larger than $\omega_k$. This result is agreement with Figure~8 of \citet{vandoorsselaere2004}. It is also consistent with Figure~5(a) of \citet{arregui2005} with their stratification parameter set to $\alpha = 0$. Although it depends on the density contrast, the actual $\omega_{\rm R}$ is around $15\%$ larger than $\omega_k$  when $l/R \approx 2$.

We turn to  the imaginary part of the frequency (Figure~\ref{fig:sin1}(b)). Visually, we see that the TTTB formula  (Equation~(\ref{eq:wi2})) provides a good approximation to the damping rate when $l/R \lesssim 0.4$. For thicker layers, the full result deviates from the linear dependence predicted in the TTTB approximation. In addition, we find that beyond the limit of thin layers the density contrast has an impact on the shape of the curves, so that the curves of $\omega_{\rm I}$ obtained for different $\rhoi / \rhoe$ do not show the same dependence with $l/R$. This dependence of the damping rate with the density contrast is not captured by the TTTB formula.  For example, the curve with $\rhoi / \rhoe = 2$ in Figure~\ref{fig:sin1}(b) saturates on a certain value and intersects the thin layer solution when $l/R \approx 1.5$. This behavior can also be seen in Figures~4 and 5 of \citet{vandoorsselaere2004}. Conversely, the rest of curves in Figure~\ref{fig:sin1}(b) corresponding to larger contrasts do not reach a saturation value and do not intersect the TTTB solution \citep[see also][Fig.~11]{vandoorsselaere2004}. In those cases the TTTB formula underestimates the actual damping rate. The largest deviation from the TTTB values takes place at $l/R \approx 1$, where the error done due to the TTTB approximation is around $25\%$. Again, these results are in good agreement with \citet{vandoorsselaere2004}.

To further check the results discussed above, we repeat the computations of the frequency and damping rate with the resistive MHD code using the same parameters as in Figure~\ref{fig:sin1}. The resistive results are overplotted in Figure~\ref{fig:sin1} with the symbol $\Diamond$. We find an excellent agreement between the ideal results obtained using the Frobenius method and the resistive MHD computations. This fact makes us confident that the analytic Frobenius method gives correct results. We do not need to use the more computationally expensive resistive MHD computations to find the kink mode frequency and damping rate for thick transitional layers.

\subsection{Eigenfunctions}

\label{sec:pertthick}

\begin{figure*}
\centering
\includegraphics[width=.99\columnwidth]{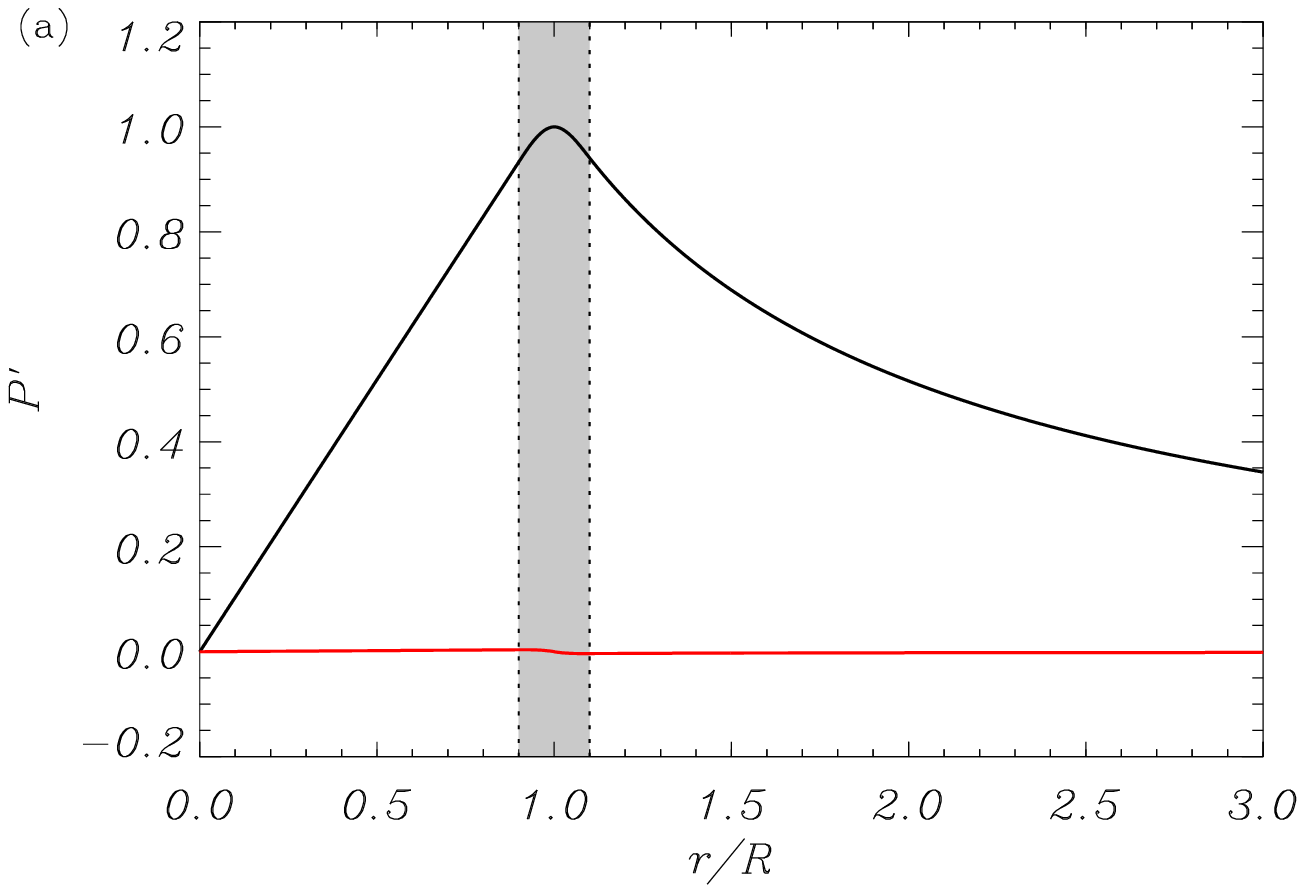}
	\includegraphics[width=.99\columnwidth]{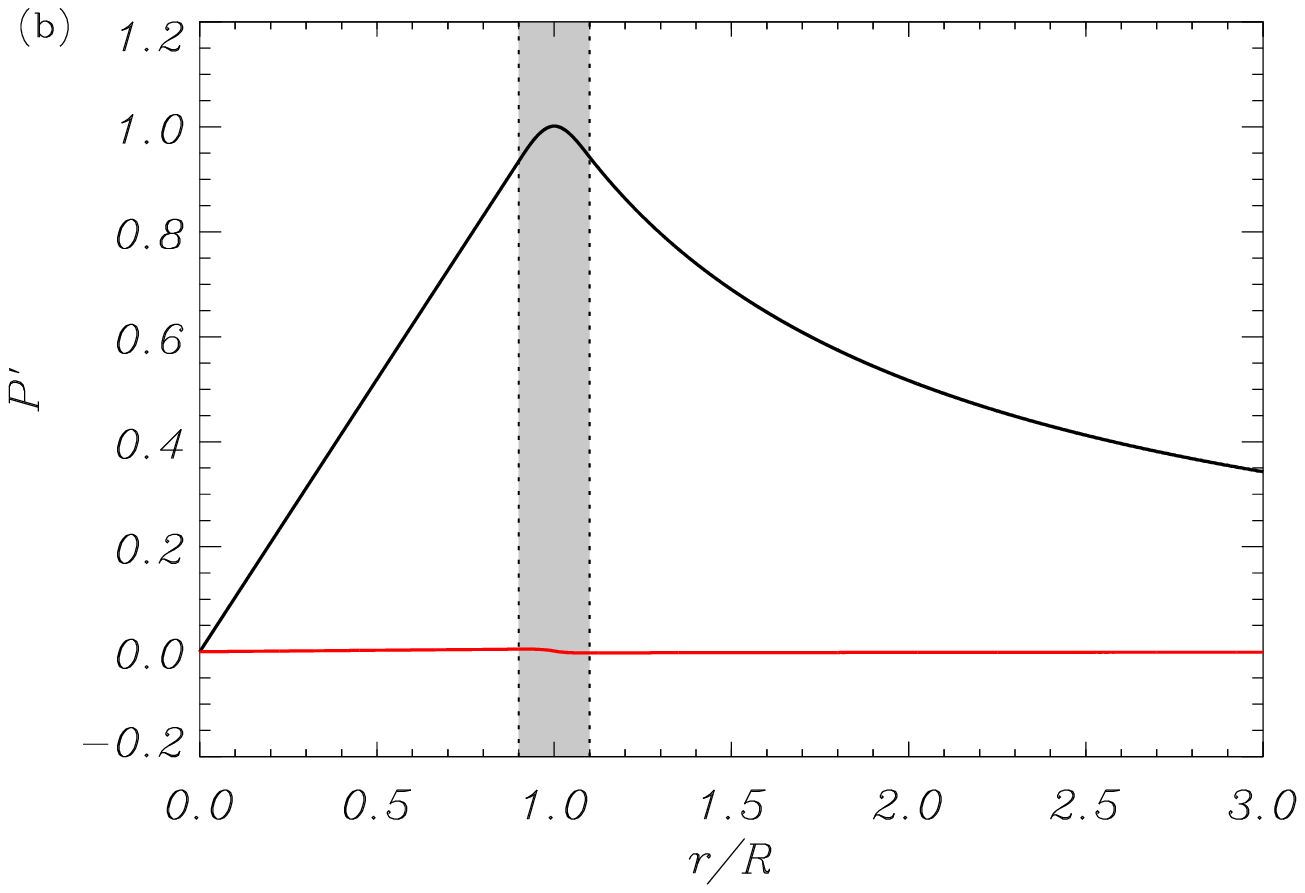}
\includegraphics[width=.99\columnwidth]{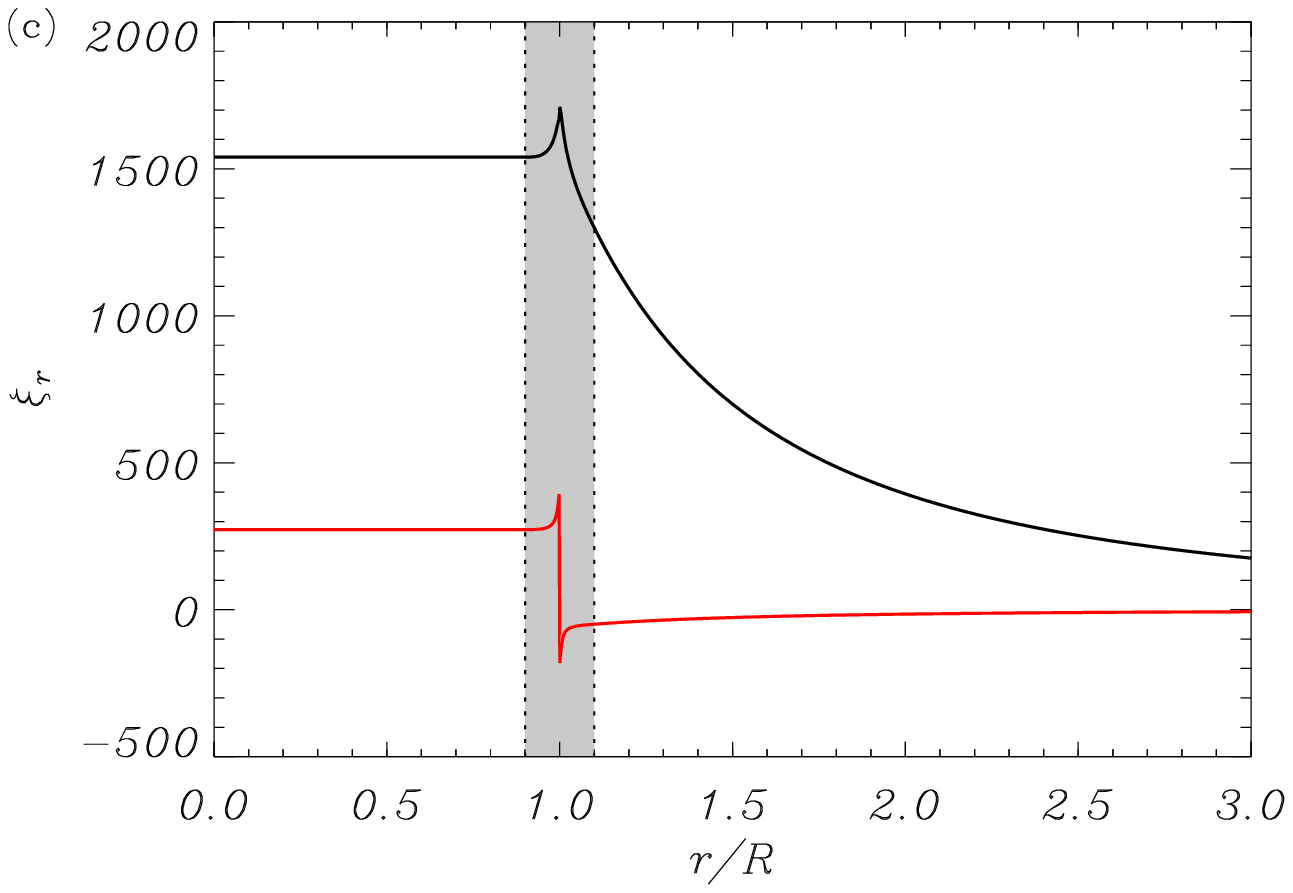}
\includegraphics[width=.99\columnwidth]{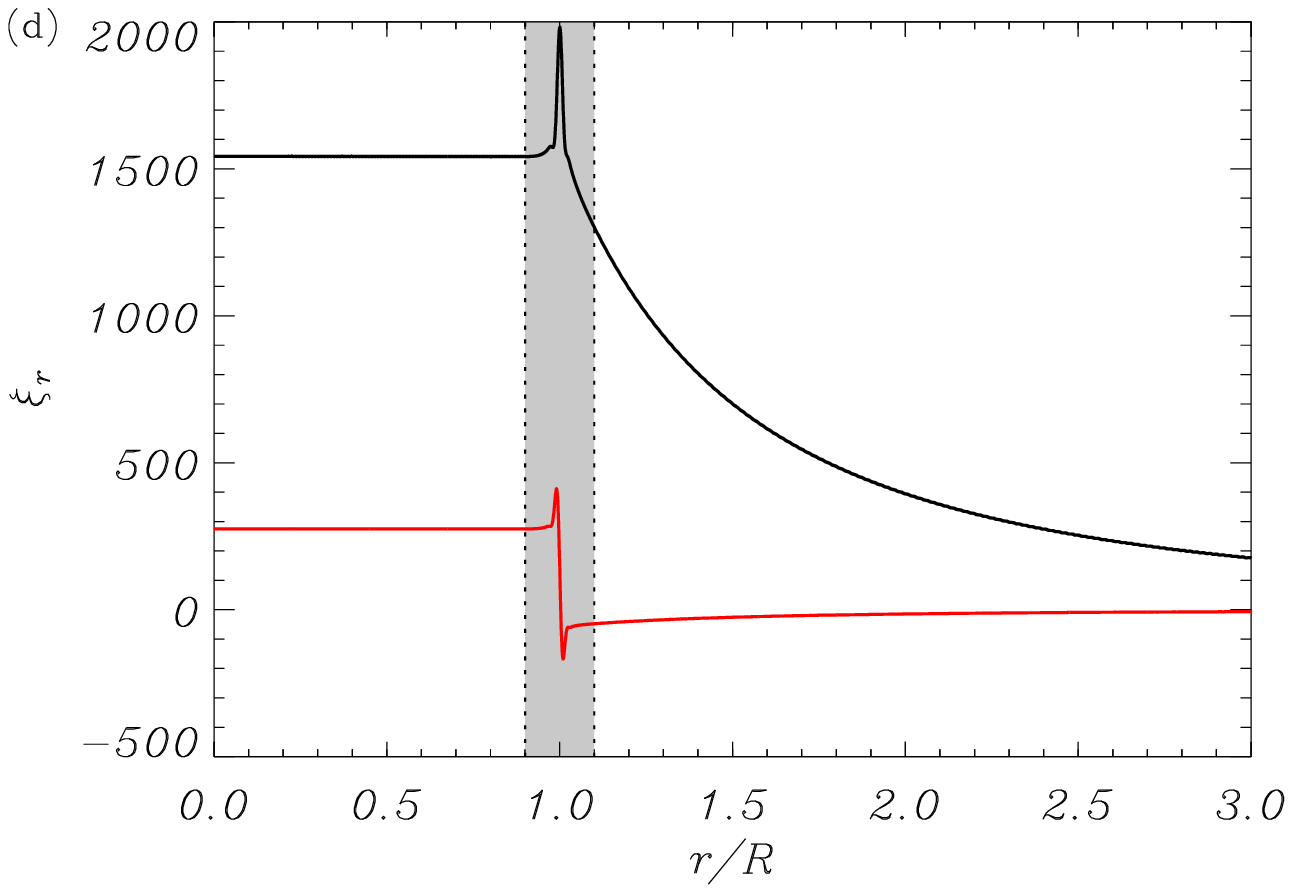}
	\includegraphics[width=.99\columnwidth]{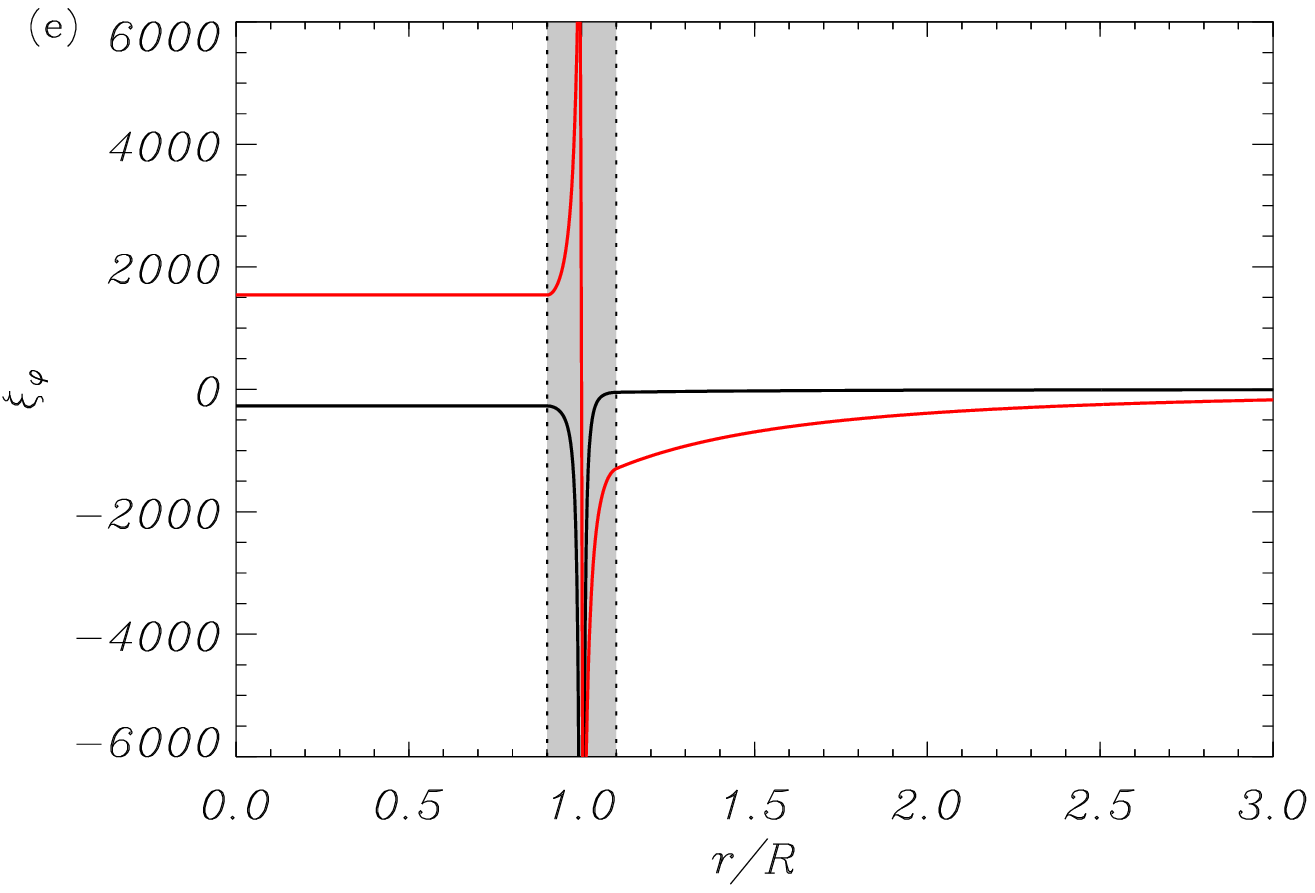}
	\includegraphics[width=.99\columnwidth]{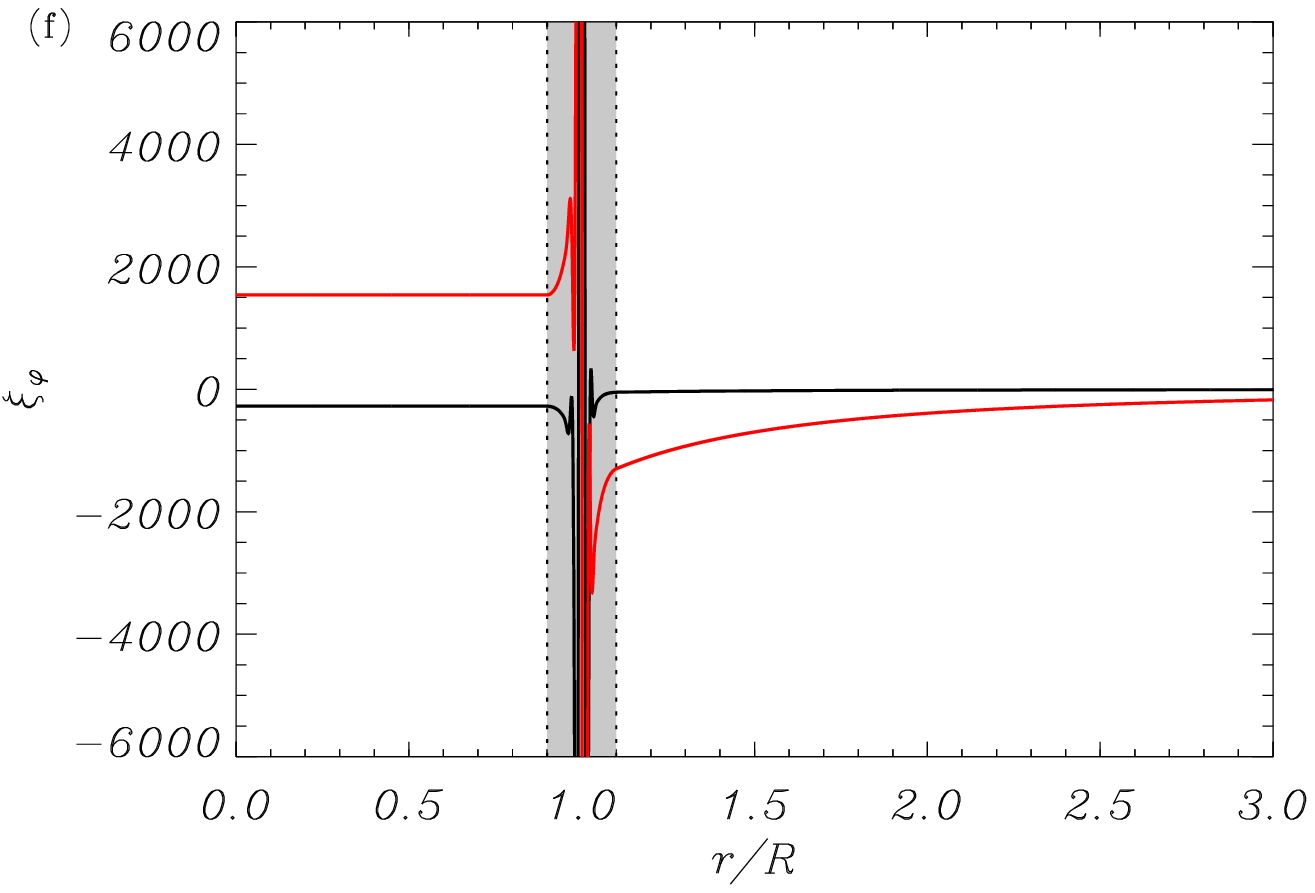}
	\caption{Kink mode eigenfunctions  $P'$ (top),  $\xi_r$ (mid), and $\xi_\varphi$ (bottom) as functions of $r/R$ for $l/R=0.2$. Left panels  are the ideal MHD eigenfunctions obtained with the Frobenius method, whereas the right panels are the resistive eigenfunctions computed with the PDE2D code. The black and red colors correspond to the real and imaginary parts of the eigenfunction, respectively. The shaded area denotes the nonuniform layer. Arbitrary units are used so that max$[{\rm Re}(P')]$=1.  We use a sinusoidal transition of density, $L/R = 100$, and $\rhoi / \rhoe = 5$.}
	\label{fig:eigen}
\end{figure*}

\begin{figure*}
\centering
\includegraphics[width=.99\columnwidth]{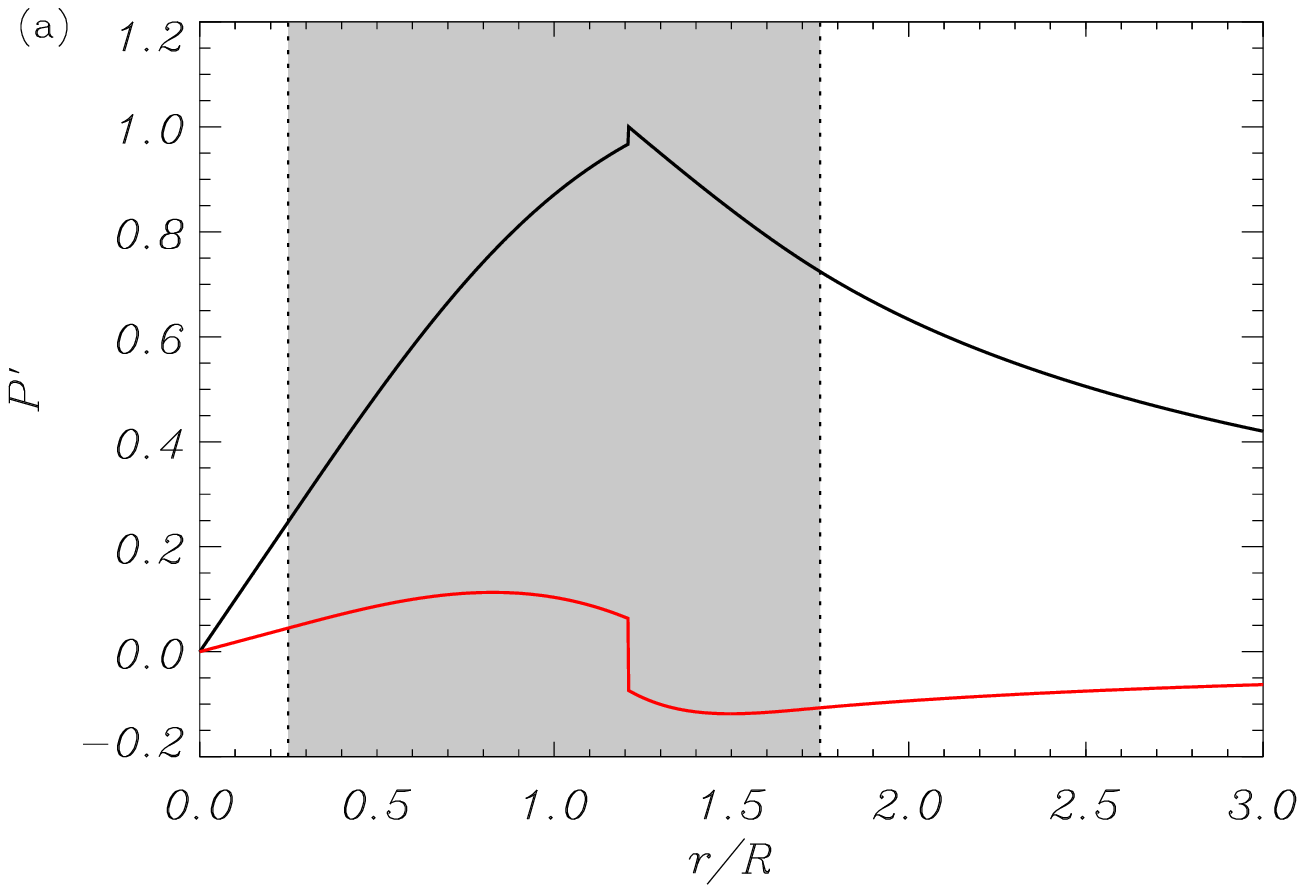}
	\includegraphics[width=.99\columnwidth]{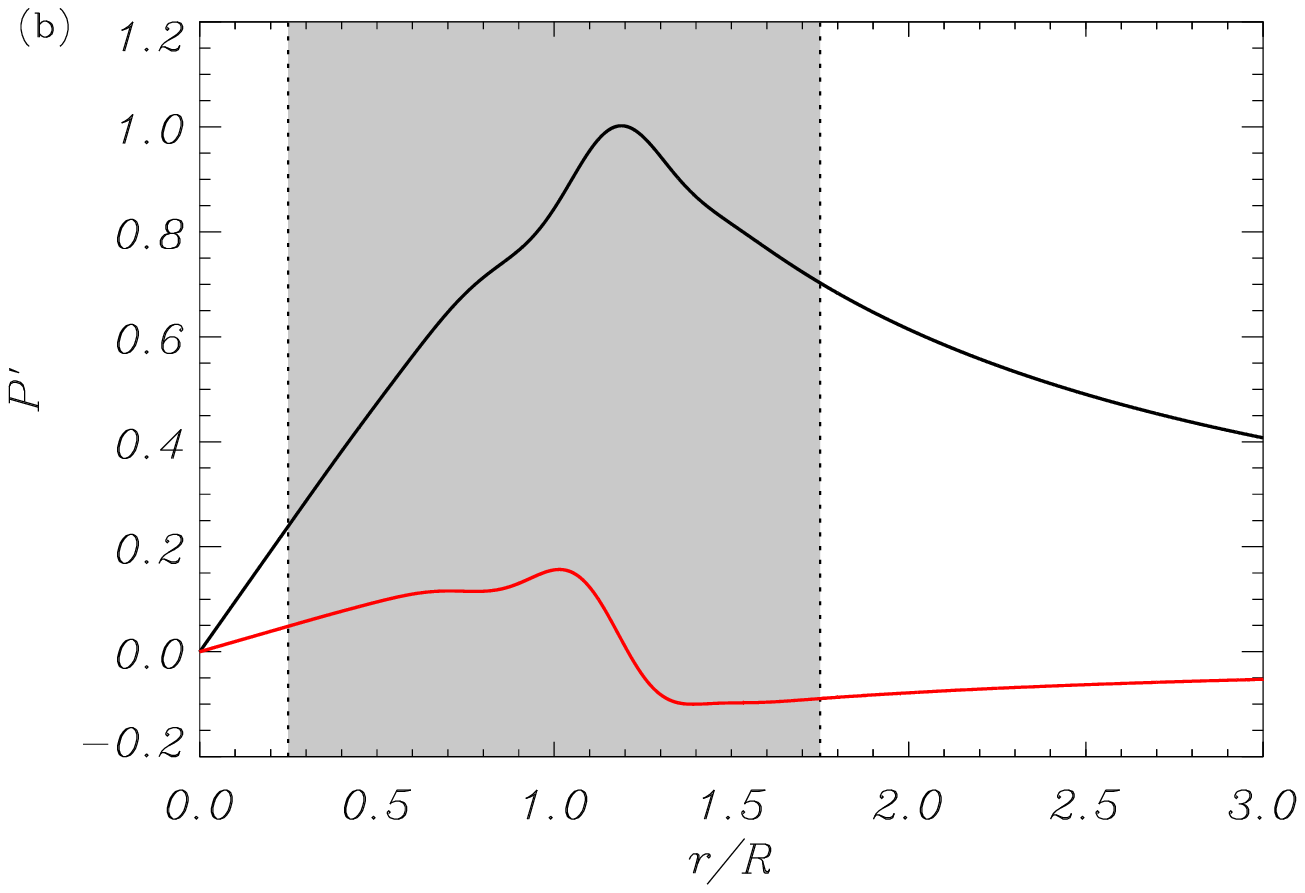}
\includegraphics[width=.99\columnwidth]{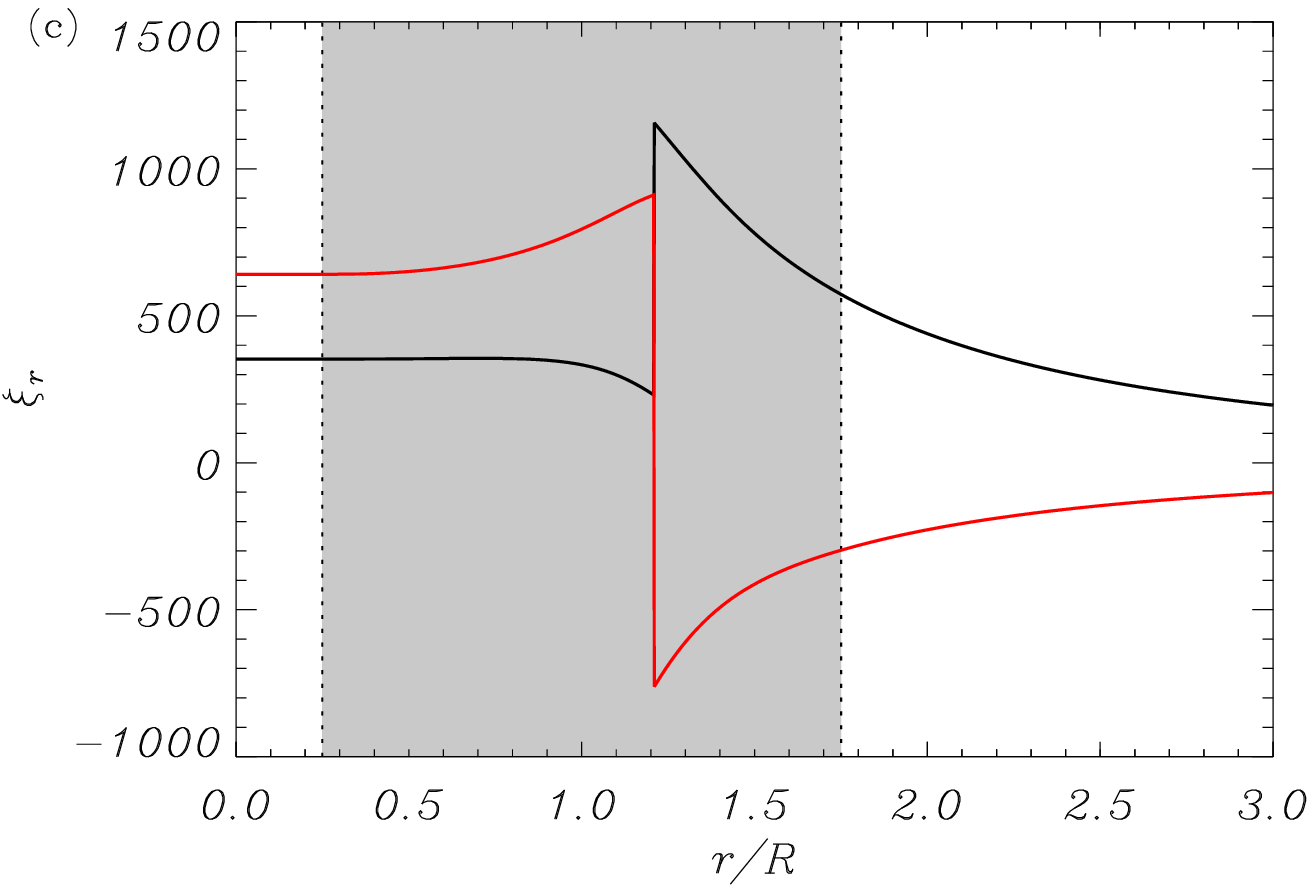}
\includegraphics[width=.99\columnwidth]{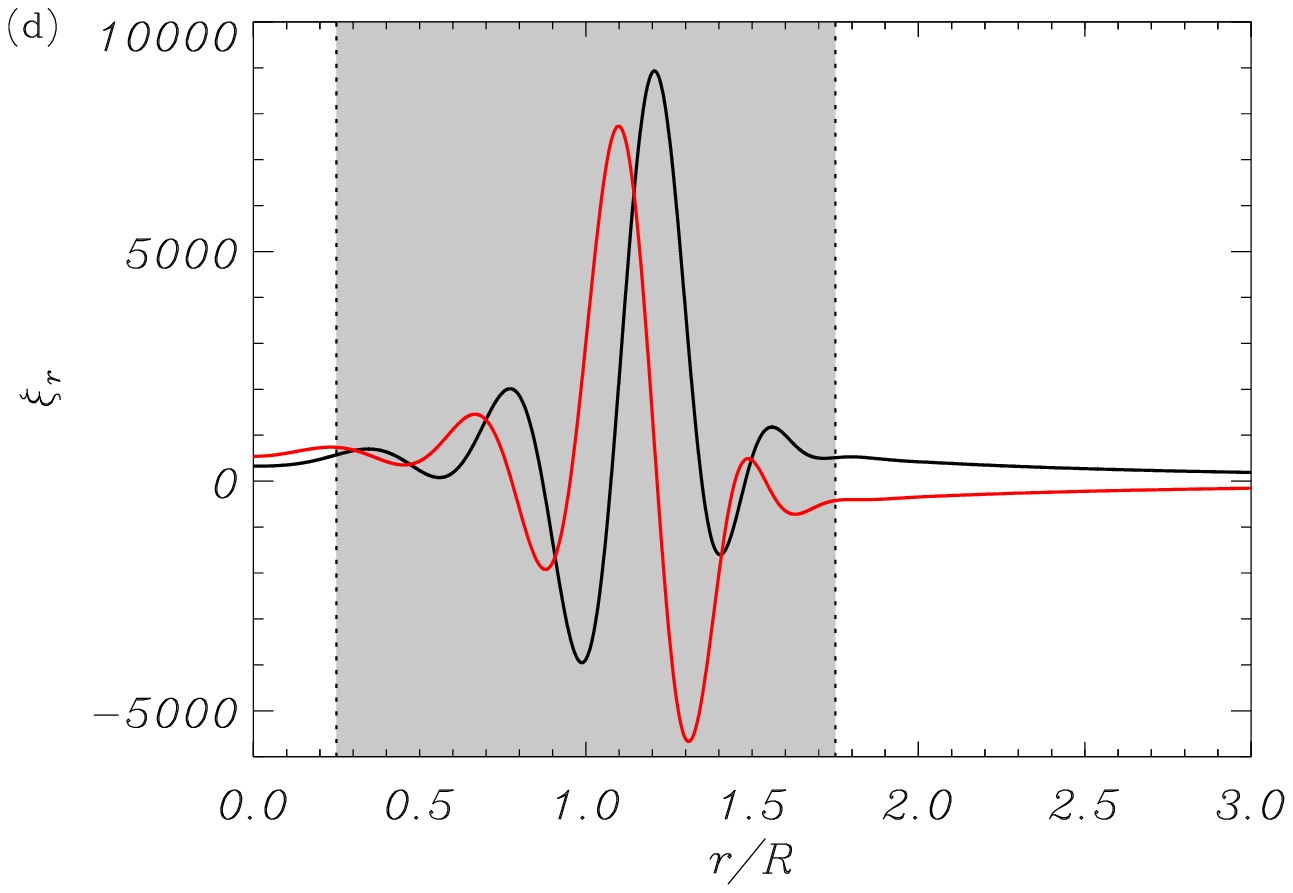}
	\includegraphics[width=.99\columnwidth]{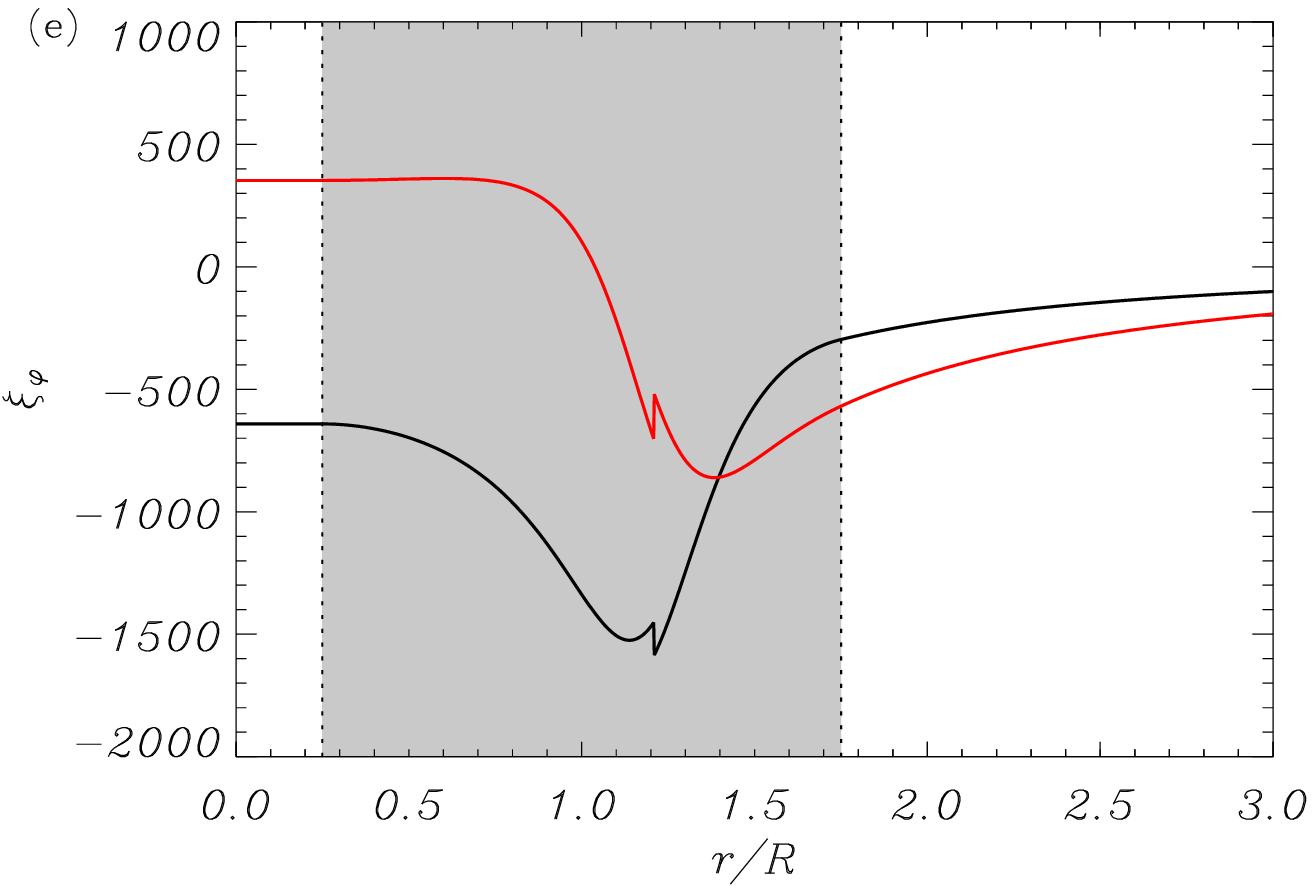}
	\includegraphics[width=.99\columnwidth]{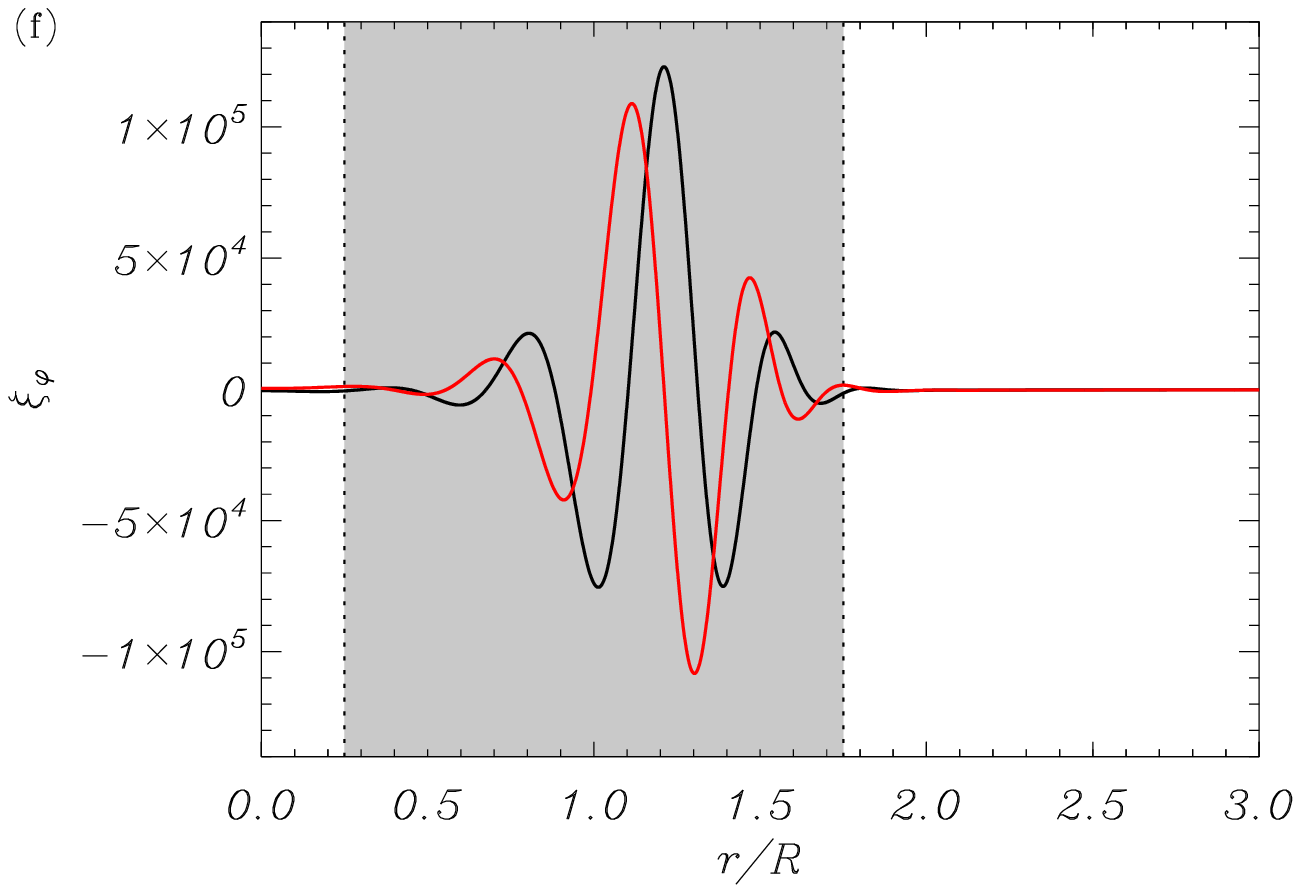}
	\caption{Same as Figure~\ref{fig:eigen} but for $l/R=1.5$.}
	\label{fig:eigen2}
\end{figure*}

\begin{figure*}
	\centering
	\includegraphics[width=.99\columnwidth]{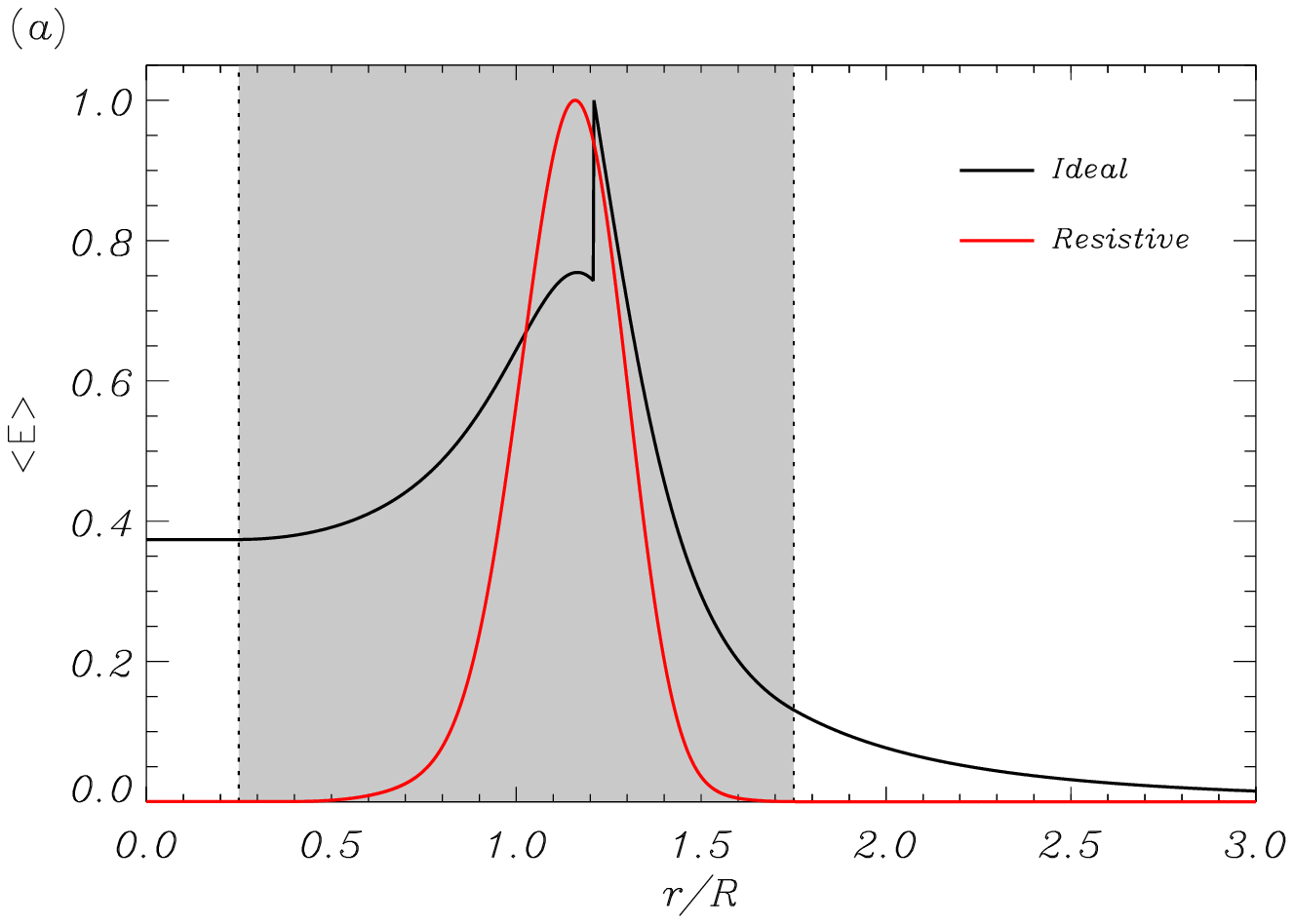}
\includegraphics[width=.99\columnwidth]{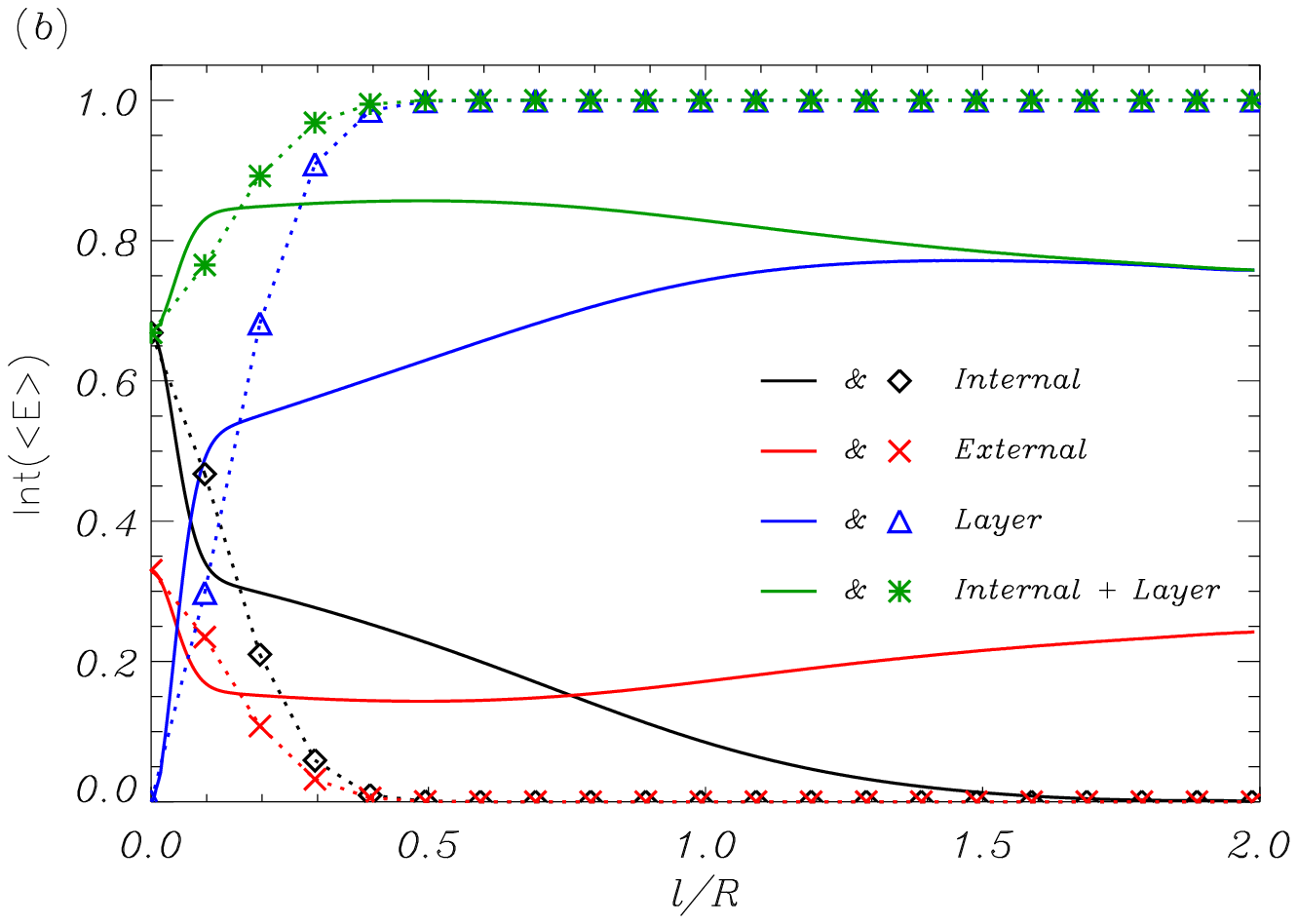}
	\caption{(a) Spatial distribution of the kink mode energy density in a tube with $l/R = 1.5$. The ideal (black) and resistive (red) results are plotted for comparison.  The shaded area denotes the nonuniform layer. The plot is normalized so that max$\left< E \right> = 1$. (b) Integrated energy in the various regions of the equilibrium flux tube as function of $l/R$. Solid lines correspond to the ideal results and symbols correspond to the resistive results. The meaning of the various colors is indicated within the Figure.  The plot is normalized with respect to the total integrated energy. In all cases we use a sinusoidal variation of density, $\rhoi / \rhoe = 5$, and $L/R = 100$. }
	\label{fig:energy}
\end{figure*}

This Subsection contains information on the eigenfunctions computed, on the one hand, with the use of the ideal MHD equations and the Frobenius method and, on the other hand, with the use of the resistive MHD equations.  The eigenfunctions were not discussed in the paper by \citet{vandoorsselaere2004}. 

We plot in Figure~\ref{fig:eigen} the ideal eigenfunctions $P'$, $\xi_r$, and $\xi_\varphi$ as functions of $r/R$ for an equilibrium flux tube with $l/R = 0.2$. The behavior of the eigenfunctions in the nonuniform layer is well described by the analytic TTTB approximations of Section~\ref{sec:pert2}. Hence,  $P'$ is almost constant, ${\rm Im}(\xi_r)$ has a logarithmic jump, and $\xi_\varphi$ varies as $\left( r - r_{\rm A} \right)^{-1}$ near the resonance position. For comparison, we also display in Figure~\ref{fig:eigen} the resistive eigenfunctions obtained with the PDE2D code. The agreement between the ideal eigenfunctions and the resistive eigenfunctions is very good. The only noticeable differences are at the center of the nonuniform layer, close to the resonance position. There, the amplitude of ${\rm Re}(\xi_r)$ is slightly larger in the resistive case than in the ideal case, and the resistive $\xi_\varphi$ displays some spatial oscillations not present in its ideal counterpart. The total pressure perturbation, $P'$, is identical in both cases.

We increase the thickness of the nonuniform layer and set $l/R = 1.5$. We are outside the range of validity of the TTTB approximation. We display in Figure~\ref{fig:eigen2} the ideal and resistive eigenfunctions corresponding to this case and notice that there are significant differences. All the ideal eigenfunctions jump at the resonance position, including $P'$. The approximation that $P'$ is constant across the resonance \citep[e.g.,][]{hollwegyang1988,sakurai1991} is not satisfied for thick layers. The constancy of $P'$ is only approximately valid when the damping is weak so that $\omega_{\rm I}^2 \ll \omega_{\rm R}^2$, as happens for thin layers. The jumps in the resistive version of $P'$ are smoothed due to the effect of resistivity. The most significant differences appear when comparing the ideal and resistive components of the Lagrangian displacement. The amplitudes of the resistive $\xi_r$ and $\xi_\varphi$ are, approximately, 10 times larger and 100 times larger, respectively, than their ideal counterparts and display spatial oscillations in the nonuniform layer instead of the jumps seen in the ideal eigenfunctions. Outside the dissipative layer the amplitudes of the resistive $\xi_r$ and $\xi_\varphi$ are negligible.

The different behavior of ideal and resistive eigenfunctions was discussed by \citet{poedts1991}. Although they considered another equilibrium, the results of \citet{poedts1991} can be directly related to the present study. \citet{poedts1991} found that the ideal normal mode frequencies are correctly recovered from the resistive eigenvalues in the limit of small resistivity. This is also the case here, because the frequency and damping rate obtained with the ideal Frobenius method fully agree with the resistive eigenvalues (see Figure~\ref{fig:sin1}). However, \citet{poedts1991} found that the resistive eigenfunctions do not converge to their ideal counterparts in the limit of small resistivity. Instead, the spatial oscillations of the resistive eigenfunctions get confined to a thinner and thinner region as $\eta$ decreases.  In the stationary case, i.e., for $\omega_{\rm I} = 0$, the spatial oscillations in the resistive eigenfunctions are confined to a region surrounding the resonance position, i.e., the dissipative layer. The thickness of the dissipative layer is measured by the quantity $\delta_{\rm A}$ \citep[see, e.g.,][]{sakurai1991}
\begin{equation}
\delta_{\rm A} = \left( \frac{\omega_{\rm R} \eta}{\left| \Delta_{\rm A} \right|} \right)^{1/3}, \label{eq:deltaa}
\end{equation}
where $\Delta_{\rm A} = \left. \frac{{\rm d}}{{\rm d}r}(\omega^2 - k_z^2 \va^2 )\right|_{\ra}$. The dissipative layer covers the interval $r\in[\ra - \epsilon\delta_{\rm A},\ra + \epsilon\delta_{\rm A}]$ with $\epsilon \approx 5$ \citep{goossens1995}. In addition, for the damped modes of interest here, i.e., for $\omega_{\rm I}< 0$, there is another length scale that determines the spatial extent of the oscillations around the resonance. This length scale due to nonstationarity, $\delta_{\rm NS}$, is given by \citep[e.g.,][]{ruderman1995,andries2003}
\begin{equation}
\delta_{\rm NS} = \left| \frac{2 \omega_{\rm R} \omega_{\rm I} }{\Delta_{\rm A}} \right|. \label{eq:deltans}
\end{equation}
Due to the effect of nonstationarity, the oscillatory domain around the resonance does not keep decreasing indefinitely when $\eta$ is decreased \citep{ruderman1995}. Instead, the width of the oscillatory domain does not decrease below a certain thickness, which is approximately given by Equation~(\ref{eq:deltans}). Once this minimal thickness is reached, the effect of decreasing $\eta$ is to produce more and more oscillations in the resonant layer. This phenomenon can be seen in Figure~2.3 of \citet{andries2003}. We observe the same behavior here. The characteristic oscillatory behavior of the resistive eigenfunctions does not disappear for vanishing resistivity so that the ideal eigenfunctions are not recovered. This is a fundamental difference between ideal and resistive eigenmodes.

The resistive results shown here can also be related to the study by \citet{vandoorsselaerepoedts2007} in a nonuniform Cartesian slab. They found that, when the nonuniform region spreads over a substantial part of the slab, the global transverse mode loses its global character and gradually becomes indistinguishable from an ordinary resistive mode of the Alfv\'en spectrum. For a similar experiment in cylindrical geometry see \citet{arregui2005}. Here, we obtain an equivalent result for the resistive kink mode. In the ideal case, however, the kink mode never loses its character as a global mode of the flux tube regardless of the thickness of the nonuniform layer.

Another important result observed in Figures~\ref{fig:eigen} and \ref{fig:eigen2} is the absence of true singularities in the ideal eigenfunctions. Instead, the eigenfunctions have finite jumps at the resonance position \citep[see also][]{stenuit1998}. As explained in Section~\ref{sec:pert2}, the finite jumps of the ideal eigenfunctions in the nonuniform layer are  caused by the logarithmic term in the Frobenius series. The physical reason for the existence of these jumps is that there is a net flux of energy toward the nonuniform layer so that $\left<  S_r  \right>$ (Equation~(\ref{eq:sr})) jumps at the resonance position \citep[see plots of the energy flux in, e.g.,][]{stenuit1999,arregui2011,goossens2013}. For thin layers, the jump of $\left<  S_r  \right>$ is given in Equation~(\ref{eq:jumpsr}). The efficiency of energy transfer towards the resonance is determined by the  jump of $\left<  S_r  \right>$. In the presence of resistivity, the expression of  $\left<  S_r  \right>$ has additional terms proportional to $\eta$, which are of minor importance compared to the ideal term and do not alter the jump of the energy flux \citep[see][]{arregui2011}. Mathematically, the absence of true singularities is a direct consequence of the fact that wave frequency, $\omega$, is complex. In turn, the singularity condition $\omega^2 = k_z^2 \va^2(\ra)$ requires $\ra$ to be a complex quantity too. Hence, there is no true singularity at $r=\ra$ because the radial coordinate, $r$, is obviously real. Instead, there is a finite logarithmic jump of the perturbations at $r={\rm Re}(\ra)$. The jump is finite as long as $\omega_{\rm I} \neq 0$ so that ${\rm Im}(\ra)\neq 0$. When $l/R \ll 1$, $\omega_{\rm I}^2 \ll \omega_{\rm R}^2$ and ${\rm Im}(\ra) \ll {\rm Re}(\ra)$, hence the behavior of the eigenfunctions is quasi-singular in thin layers (Figure~\ref{fig:eigen}), although we stress that there is no true singularity even in this case. Conversely, for thick layers $\omega_{\rm I}^2$ and $\omega_{\rm R}^2$ are of the same order and there is no hint of singularity in the eigenfunctions (Figure~\ref{fig:eigen2}). The absence of singularities in the eigenfunctions is a result also found in time-dependent simulations, where the wave perturbations remain finite around the resonance position \citep[see, e.g.,][]{terradas2006,soler2011strat,pascoe2013}.

\subsection{Energy distribution}

The distinct form of ideal and resistive eigenfunctions has important repercussions for the computation of the wave energy distribution. The time-averaged total energy density, $\left< E \right>$, is \citep[e.g.,][]{walker2005}
\begin{equation}
\left< E \right> = \frac{1}{2}  \left( \rho {\bf v} \cdot {\bf v}^* + \frac{1}{\mu} {\bf b} \cdot {\bf b}^* \right), \label{eq:energy}
\end{equation} 
where ${\bf v} = - i \omega \xii$ is the velocity perturbation. We use the eigenfunctions for $l/R=1.5$ displayed in Figure~\ref{fig:eigen2} to compute the spatial distribution of $\left< E \right>$. This is shown in Figure~\ref{fig:energy}(a). As expected, the ideal and resistive results differ. In the resistive case, the energy  is essentially confined near the vicinity of the resonance, whereas the amount of energy in the rest of the equilibrium is negligible. The energy spatial distribution depends upon $\eta$, so that the smaller $\eta$, the more confined is the energy around the resonance position. Conversely, in the ideal case energy spreads over the whole nonuniform layer, and the amount of energy in the internal and external regions is not negligible. 

We study how the energy is distributed in the flux tube when $l/R$ varies from $l/R = 0$ to $l/R=2$ by computing the integrated energy density,
\begin{equation}
{\rm Int}\left( \left<E\right> \right) = \int_0^\infty \left< E \right>  r {\rm d}r. \label{eq:integratedE}
\end{equation}
Figure~\ref{fig:energy}(b) displays the integrated energy density in the various regions of the equilibrium as function of $l/R$.  Both ideal and resistive results agree when $l/R=0$. Around $65\%$ of the wave energy is in the internal plasma and around $35\%$ in the external plasma. The differences between ideal and resistive results arise when $l/R$ increases.  In the case of the resistive results, all the energy goes to the nonuniform layer when $l/R \gtrsim 0.5$. Almost no energy is left in the internal and external regions in the resistive case.  In the ideal case, most of the energy is in the nonuniform layer too, but the amount of energy in the internal and external plasmas is not negligible even when the transitional layer is very thick. For a fully nonuniform tube, i.e., $l/R = 2$, around $25\%$ of the energy remains located in the external medium in the ideal case. 

Recently, \citet{goossens2013} discussed the energy distribution of kink waves in nonuniform flux tubes. Since they used results from resistive eigenvalues, their conclusion was that almost all the energy is concentrated in the nonuniform layer. Based on this result, they proposed a simple formula to compute the total energy carried by kink waves in the solar atmospheric flux tubes \citep[Equation~(48)]{goossens2013}. In view of the  results shown here, the conclusions of  \citet{goossens2013} should be reconsidered for the case of the energy distribution of ideal kink waves.

\section{ROLE OF THE DENSITY PROFILE}  
\label{sec:profile}  
  
In the  previous Section~\ref{sec:thick}, we used a sinusoidal variation of  density in the transitional layer. Here, we determine the impact of using other density profiles. In addition to the sinusoidal variation (Equation~(\ref{eq:sin})) we consider a linear profile,
\begin{equation}
\rho_{\rm tr}(r) = \rhoi - \frac{\rhoi - \rhoe}{l}\left( r - R + \frac{l}{2} \right),
\end{equation} 
and a parabolic profile,
\begin{equation}
\rho_{\rm tr}(r) = \rhoi - \frac{\rhoi-\rhoe}{l^2} \left( r - R + \frac{l}{2} \right)^2.
\end{equation}  
We have chosen these two profiles  for the following reason. The numerical factor $F$ is the only effect of the specific density variation that remains in the TTTB formula for the damping rate (Equation~(\ref{eq:wi2})). This is based on the assumption that $\ra \approx R$, so that the formula to compute the factor $F$ is
\begin{equation}
F = \frac{4}{\pi^2} \frac{l}{\rhoi - \rhoe} \left| \frac{\der \rho}{ \der r} \right|_R.
\end{equation}
Hence, $F=2/\pi$ for the sinusoidal profile and $F=4/\pi^2$ for both linear and parabolic profiles. Using these values of $F$ in Equation~(\ref{eq:wi2}), the same damping rate is predicted for both linear and parabolic profiles. On the other hand, the form of the density profile has no impact at all on the real part of the frequency according to Equation~(\ref{eq:wrl}). Our fist aim here is to check the validity of these predictions.

We must point out that, almost certainly, the true shape of the transitional layer in coronal flux tubes is not any of the three profiles selected here. The second aim of this Section is to determine whether our ignorance about this shape is relevant or, on the contrary, it is unimportant for the behavior of the kink mode.

\begin{figure*}
	\centering
	\includegraphics[width=.99\columnwidth]{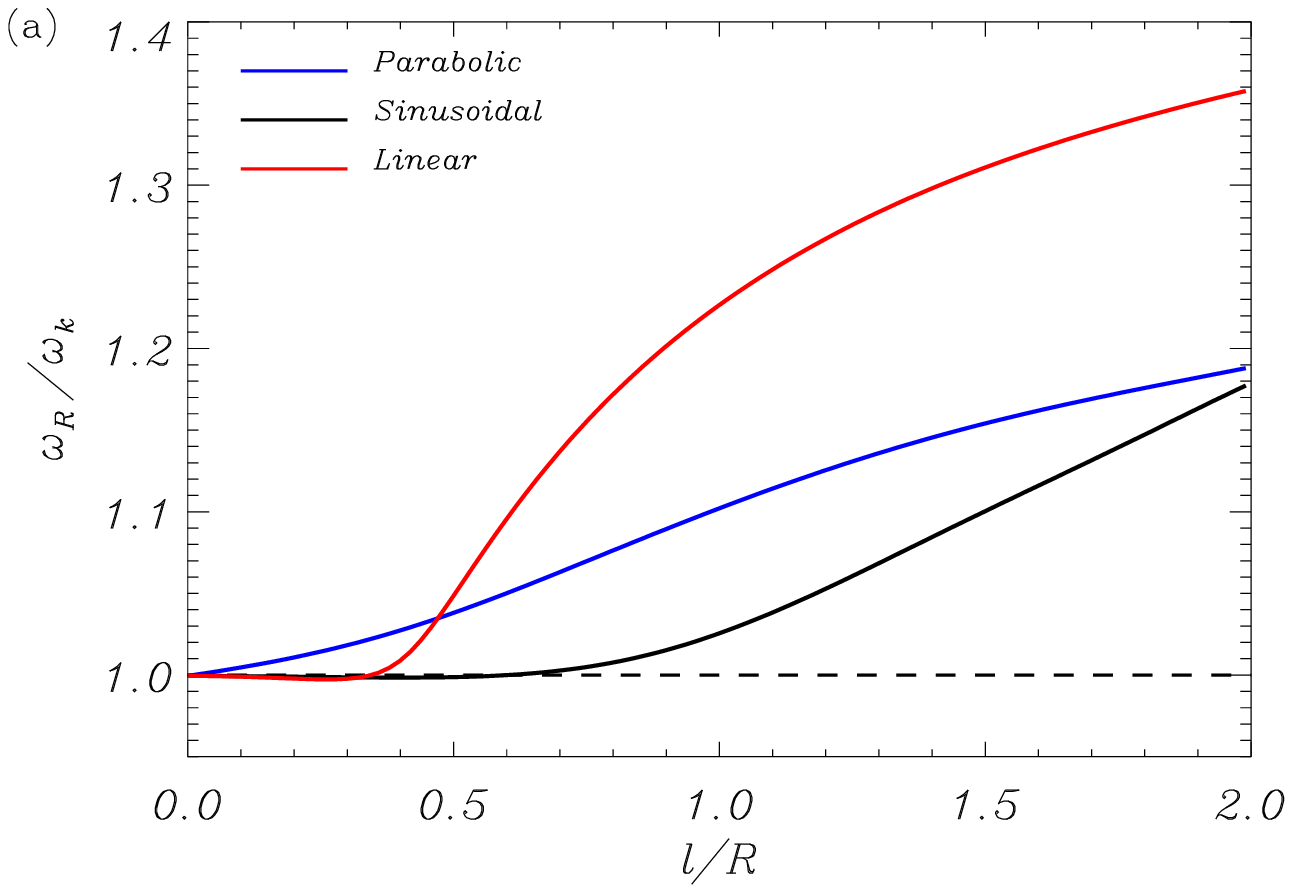}
\includegraphics[width=.99\columnwidth]{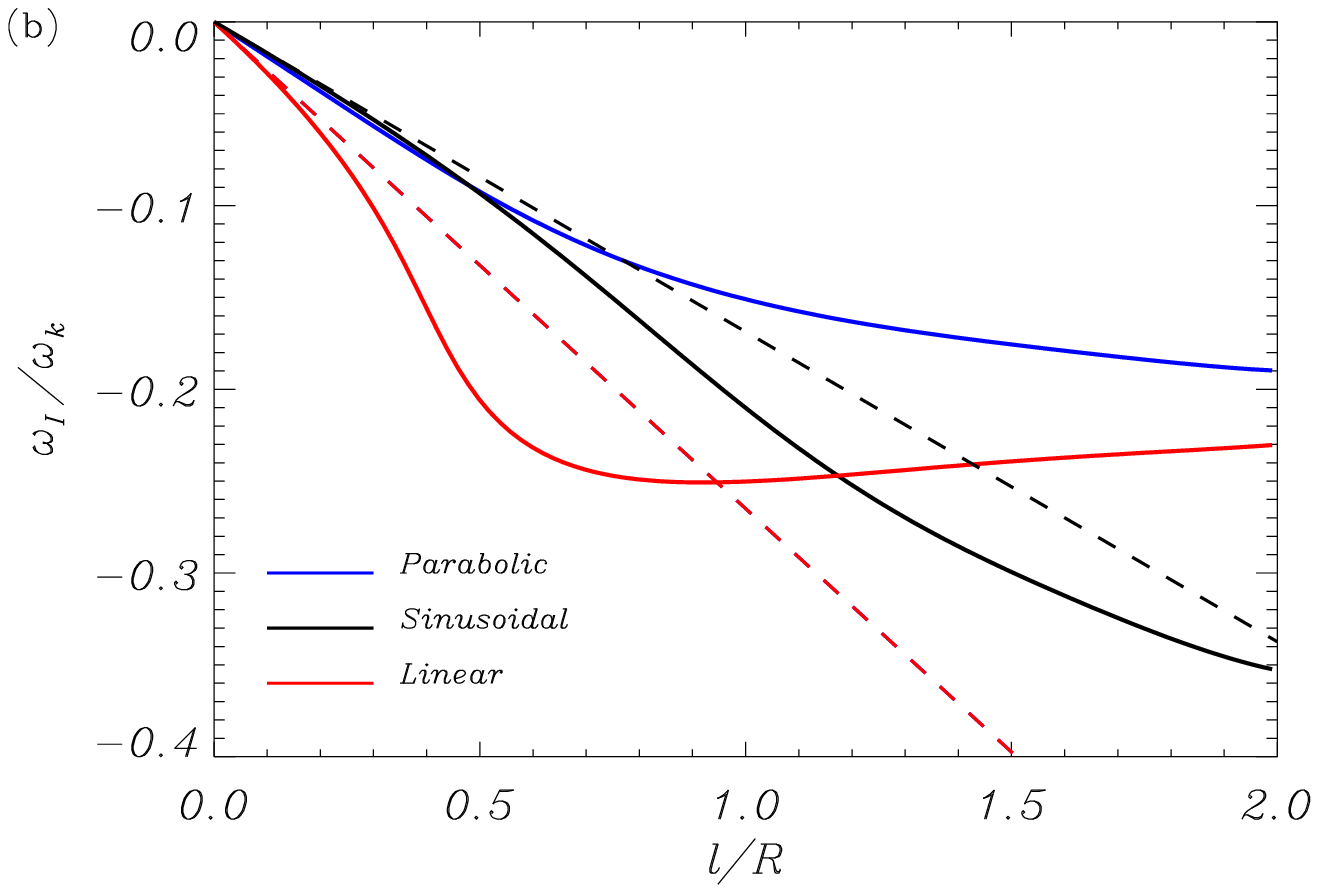}
	\caption{(a) Real part and (b) imaginary part of the kink mode frequency versus $l/R$. The line color denotes the density profile considered in the nonuniform layer (indicated within the figures). The dashed lines are the TTTB analytic results (Equations~(\ref{eq:wrl}) and (\ref{eq:wi2})). The red dashed line applies to both linear and parabolic profiles. In all cases we  use $\rhoi / \rhoe = 5$ and $L/R = 100$. }
	\label{fig:compara}
\end{figure*}

\subsection{Effect on the frequency and damping rate}

Figure~\ref{fig:compara} shows the real and imaginary parts of the fundamental kink mode frequency versus $l/R$ for the three considered density profiles with $L/R=100$ and $\rhoi/\rhoe = 5$. At first sight, we notice striking differences between the various curves. 

We start by analyzing the behavior of $\omega_{\rm R}$ (Figure~\ref{fig:compara}(a)). We find that the density profile affects the value of $\omega_{\rm R}$ beyond the thin layer limit. We compare the results of the three profiles.
\begin{enumerate}
\item The result for the sinusoidal profile is the solution that deviates the least from the TT approximation. It deviates from $\omega_{\rm R} \approx \omega_k$ around $l/R \approx 0.7$, which is a relatively thick layer. The actual $\omega_{\rm R}$ is $18\%$ larger than $\omega_k$  when $l/R \approx 2$.
\item The value of $\omega_{\rm R}$  for the linear profile deviates from $\omega_k$ around $l/R \approx 0.4$, which is a thinner layer than for the sinusoidal profile. Later, it is the solution that departs the most from the TT approximation. In this case, the actual $\omega_{\rm R}$ is $35\%$ larger than $\omega_k$  when $l/R \approx 2$.
\item The solution for the parabolic profile  deviates immediately from the TT approximation when $l/R$ increases from zero, although $\omega_{\rm R}$ does not increase as much as for the linear profile. When $l/R \approx 2$, the actual $\omega_{\rm R}$ is $19\%$ larger than $\omega_k$. 
\end{enumerate}

We turn to the damping rate (Figure~\ref{fig:compara}(b)). Again, the results for the three profiles are significantly different. Two relevant findings should be stressed.
\begin{enumerate}
\item For small $l/R$ the damping rate for the parabolic profile is closer to the solution for the sinusoidal profile than to that for the linear profile, although the TTTB formula predicts the same damping rate for both  linear and parabolic profiles. The reason for this discrepancy is that $\ra \approx R$ is a bad approximation for the parabolic profile. For the parabolic dependence, the density is not symmetric with respect to $r=R$ and the actual resonance position is not located at the center of the transitional layer. Consequently, using $F=4/\pi^2$ for the parabolic profile causes the TTTB formula to overestimate the actual damping rate. Accidentally, using $F = 2/\pi$ as for sinusoidal profile provides an approximate damping rate closer to the actual value. 
\item The damping rate for the linear profile is nonmonotonic. For the parameters used in Figure~\ref{fig:compara}, there is a  turning point at $l/R \approx 0.9$. As a consequence, we find that the behavior for $l/R \gtrsim 0.9$ is the opposite one to the prediction of the TTTB formula, i.e.,  the thicker the layer, the weaker the damping. In addition, an intersection of the actual damping rate with the TTTB approximate value occurs for thick layers regardless the density contrast.  The presence of the turning point has also the consequence that  two different values of $l/R$ produce the same damping rate.
\end{enumerate}

The results for the linear density profile shown here can be compared with the study by \citet{tatsuno1998}. These authors investigated the resonant damping of surface Alfv\'en waves in a Cartesian slab in the limit of propagation nearly perpendicular to the magnetic field. They took a linear variation of density in a nonuniform layer surrounding the slab dense core. Although the Cartesian model used by \citet{tatsuno1998} is different from the cylindrical waveguide used here, the present results for the behavior of the kink mode frequency and damping rate are similar to those found by \citet{tatsuno1998}, as can be seen by comparing their Figure~2 with our Figure~\ref{fig:compara}.

An explanation for the different behavior of  the damping rate for the three  profiles  can be found in the dependence of the jump of the radial component of the energy flux with $l/R$. We compute the radial energy flux at the resonance, $F_r$, as \citep[see, e.g.,][]{wright1994,arregui2011}
\begin{equation}
F_r = \ra  \left| \left[ \left< S_r \right> \right]_{\ra} \right|,
\end{equation}
where $\left[ \left< S_r \right> \right]_{\ra}$ denotes the jump of $\left< S_r \right> $ at the resonance. Following \citet{wright1994}, see also \citet{arregui2011}, an alternative way to obtain the kink mode damping rate (in absolute value) is computing the ratio of $F_r$ to the total integrated energy, ${\rm Int}\left( \left<E\right> \right)$, namely
\begin{equation}
\left| \omega_{\rm I} \right|  = \frac{F_r}{{\rm Int}\left( \left<E\right> \right)}.
\end{equation}
This result is displayed in Figure~\ref{fig:comparaE}(a), where it is compared to the actual $\omega_{\rm I}$ plotted in Figure~\ref{fig:compara}(b). An excellent agreement between both results is obtained. The form of the density profile affects the evolution $F_r$ when $l/R$ increases. In the case of the linear profile, we find that for thick layers  $F_r$ gets smaller  when $l/R$ increases. As a consequence, the efficiency of resonant damping becomes lower when the thickness of the layer grows. Conversely, for the parabolic and sinusoidal profiles $F_r$ gets larger when $l/R$ increases.

In summary, we find that beyond the limiting case of thin nonuniform layers, the specific form of the density profile strongly influences the kink mode frequency and damping rate.  Among the three profiles used here, the sinusoidal variation is the profile for which the extrapolation of the TTTB approximation works the best.  For other density profiles different form the sinusoidal one, the error done by the TTTB approximation when used beyond its range of applicability can be much larger depending on the model parameters. A detailed study of the error done due to the TTTB approximation will be presented in the forthcoming second part of this work.

\subsection{Effect on the energy distribution}

\begin{figure*}
	\centering
		\includegraphics[width=.99\columnwidth]{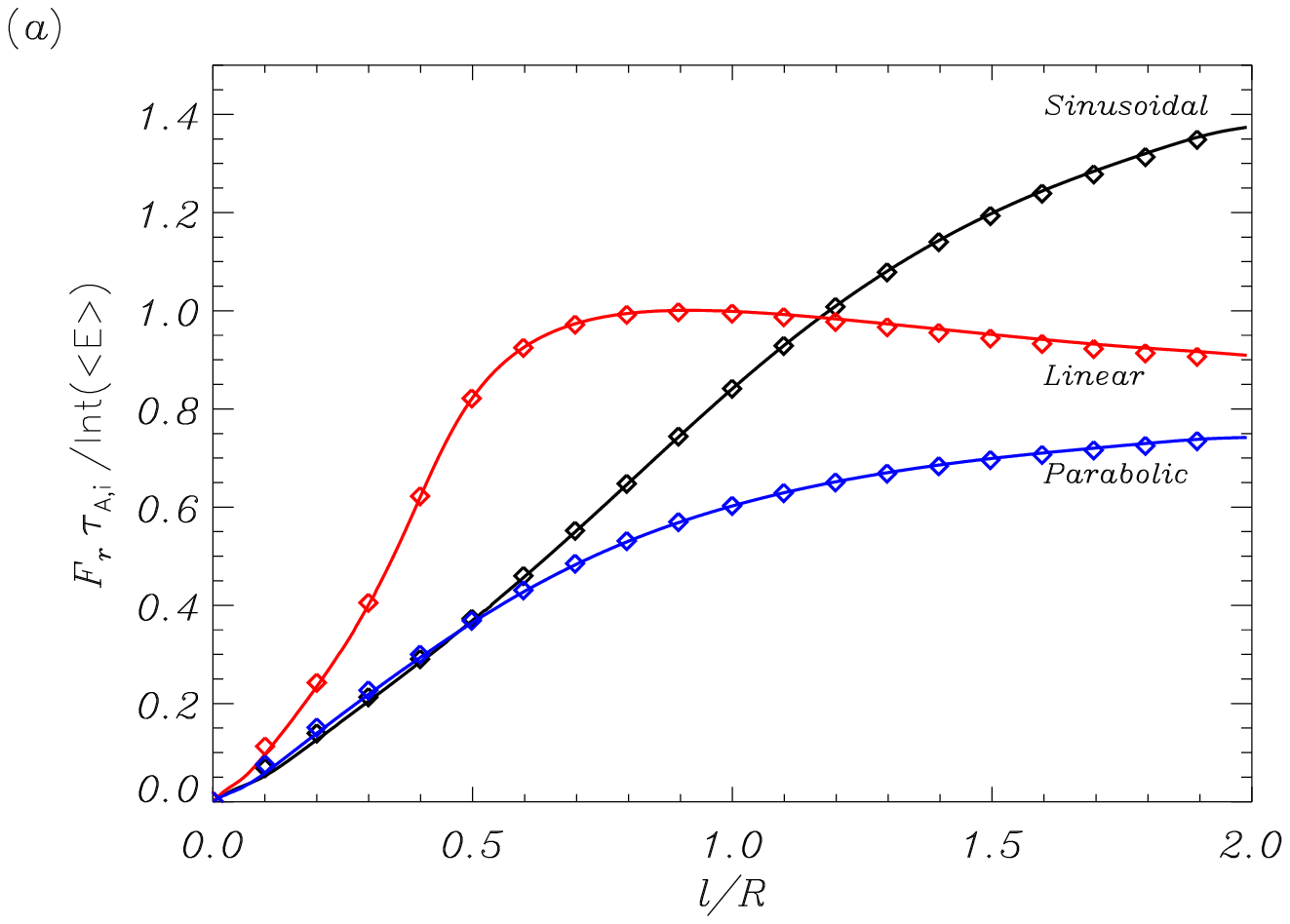}
	\includegraphics[width=.99\columnwidth]{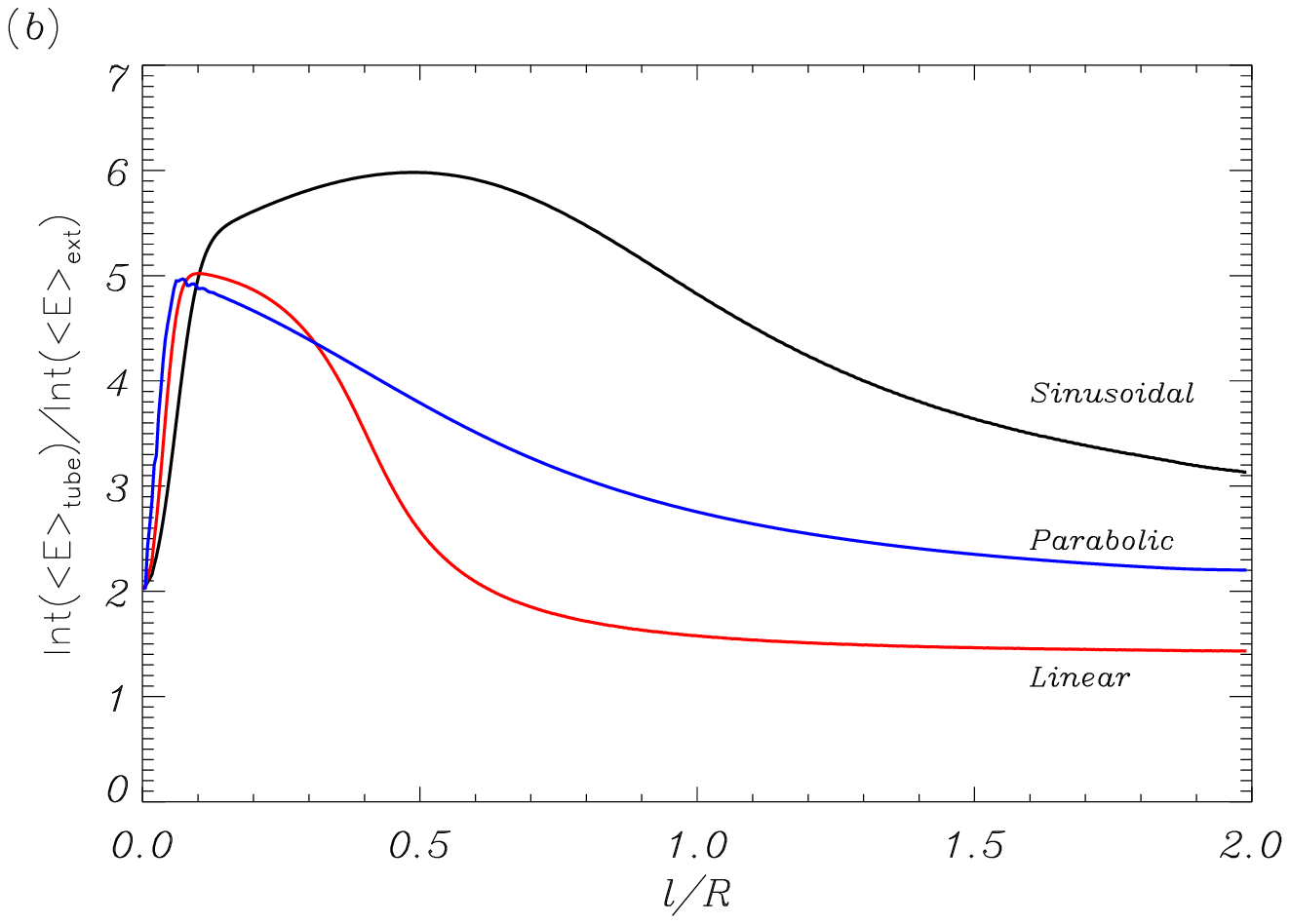}
	\caption{(a) Damping rate as function of $l/R$. Solid lines are  the ratio of the radial energy flux at the resonance, $F_r$, to the integrated total energy.   Symbols are the absolute value of $\omega_{\rm I}$ plotted in Figure~\ref{fig:compara}(b). In both cases, the damping rate is given in units of the internal Alfv\'en travel time, $\tau_{\rm A,i} = L/\vai$. (b) Ratio of the integrated energy inside the flux tube (internal medium + nonuniform layer) to the integrated energy in the external plasma as function of $l/R$. In both panels, the line color denotes the density profile considered in the nonuniform layer. In all cases we use $\rhoi / \rhoe = 5$ and $L/R = 100$.}
	\label{fig:comparaE}
\end{figure*}

Finally, we determine the effect of the transverse density profile on the wave energy distribution. To this end, we compute ${\rm Int}\left( \left<E\right> \right)$ in the various regions of the equilibrium as function of $l/R$. We use the ideal eigenfunctions obtained with the Frobenius method. Figure~\ref{fig:comparaE}(b) shows the ratio of the integrated energy inside the flux tube, i.e., the integrated energy in the internal homogeneous medium plus the integrated energy in the nonuniform layer, to the integrated energy in the external plasma. This ratio informs us about the amount of energy confined within the flux tube compared to the amount of energy located in the external medium. When $l/R=0$, the energy inside the flux tube is about twice the energy in the external medium. Again, we find important differences between the results obtained with the three density profiles when $l/R$ increases.  

Among the three profiles, the sinusoidal one is the profile for which the confinement of energy inside the flux tube is largest, while the linear profile is the profile that produces the poorest confinement of energy. When $l/R = 2$, the amount of energy inside the tube is about 3 times larger than outside for the sinusoidal profile, whereas for the linear profile the energy inside the tube is only 1.5 times larger than the external energy, approximately. The result for the parabolic profile is in between those of the linear and sinusoidal profiles.

As for the frequency and damping rate, the results here show a strong dependence of the energy distribution on the specific density variation in the nonuniform layer.  For comparison, we have repeated the computations of Figure~\ref{fig:comparaE}(b) but using the resistive eigenfunctions (not shown here). In the resistive case, there are no differences between the various profiles because all the energy quickly goes to the dissipative layer when $l/R$ increases from zero, so that the amount of energy in the external plasma becomes negligible regardless of the density profile (see Figure~\ref{fig:energy}(b)). 
  
\section{DISCUSSION}  

\label{sec:conclusion}  
  
In this article, we developed an analytic technique to compute the dispersion relation and the eigenfunctions of transverse MHD waves in pressureless cylindrical flux tubes with nonuniform transitional layers of arbitrary thickness. The method allows to consider an arbitrary spatial variation of density in the nonuniform layer. Unlike previous works that considered thick transitional layers \citep[see, e.g.,][]{vandoorsselaere2004,arregui2005}, the technique does not rely on the use of resistive MHD computations. We consider the linearized ideal MHD equations and use the Method of Frobenius to express the solution for the total pressure perturbation in the nonuniform layer as a combination of a singular and a regular series around the Alfv\'en resonance position \citep[see, e.g.,][]{zhu1988,hollweg1990a,hollweg1990b,wright1994,cally2010}. Specific results for kink modes were produced as an application of the technique. We compared the ideal results obtained with the Frobenius method with the fully numerical solution of the resistive eigenvalue problem in the limit of small resistivity. 

We find that the frequency and resonant damping rate of kink waves is the same in both ideal and resistive cases. The comparison of the results obtained with different density profiles in the nonuniform layer revealed that the specific form of the density variation affects both the frequency and the damping rate  beyond the limit of thin layers. In particular, the accuracy of the TTTB approximation for the ratio of the damping time to the period is very sensitive to the density profile.  The previous papers that studied the effect of thick layers \citep{vandoorsselaere2004,arregui2005} restricted themselves to a sinusoidal variation and did not consider other density profiles. The error done with the TTTB formula of $\tau_{\rm D}/P$ might be larger than 25\%, which was estimated by \citet{vandoorsselaere2004} for the specific case of a sinusoidal variation. This will be explored in detail in the second part of this investigation.

The ideal MHD eigenfunctions are slightly different for thin nonuniform layers but strongly different for thick layers from their resistive MHD counterparts.  The ideal eigenfunctions show finite jumps at the Alfv\'en resonance position as a consequence of the net energy inflow into the resonance from both sides.  In the resistive case, the jumps are replaced by spatial oscillations of large amplitude, which are very localized around the resonance position, so that the kink mode loses its global character and becomes indistinguishable from a resistive Alfv\'en mode \citep{vandoorsselaerepoedts2007}. This raises a fundamental theoretical problem about the correspondence of ideal MHD solutions and resistive MHD solutions \citep[see, e.g.,][]{poedts1991,ruderman1995,andries2003}.

Importantly, we stress  the absence of  singularities in the   eigenfunctions. From the mathematical point of view, the absence of  singularities is a direct consequence of the fact that that normal mode frequency, $\omega$, is complex. Equivalently, singularities are also absent in the case of propagating waves, because $k_z$ is then complex. When either $\omega$ or $k_z$ are complex, the resonance condition $\omega^2 = k_z^2 \va^2\left( \ra \right)$ implies that the resonance position, $\ra$, is in the complex plane. For a singularity to be present, $\ra$ has to be on the real axis. This would be the case, for example, of a line-tied flux tube that is externally forced to oscillate at a certain frequency, so that both $\omega$ and $k_z$ are fixed real quantities and, therefore, $\ra$ is real. For damped normal modes, however, singularities do not occur, although we  can call the position $r={\rm Re}\left( \ra \right)$ a quasi-singularity.

The different form of the  eigenfunctions affects the spatial distribution of energy carried by the waves. In the resistive case, the wave energy distribution is essentially confined in the vicinity of the resonance position regardless of the transverse density profile. In the ideal case, the energy spreads over the whole flux tube, and  the amount of energy that is confined within the waveguide or is located in the external medium depends on the transverse density profile.

 The question of what description, i.e., ideal or resistive, is the best representation for the observed waves in the solar atmosphere is open for discussion. This question is not only of obvious theoretical importance but also of practical importance for the interpretation of the energy content and distribution in kink waves \citep{goossens2013}. In the solar atmosphere magnetic resistivity is very small but it is definitely nonzero. The problem of how to reconcile the observed transverse global oscillations with the resistive eigenmodes is a  serious challenge for future theoretical studies.

Another important issue is how to relate the results from normal modes with those of the time-dependent solution. Normal modes provide consistent values of period and damping time when compared with those obtained from the time-dependent solution \citep[e.g.,][]{terradas2006}. However, time-dependent simulations show that the dynamics in the vicinity of the resonance after several periods does not correspond to the behavior expected from the normal mode. As the global oscillation damps, the energy fed into the inhomogeneous layer is used to generate small-scale motions near the resonance position. The amplitude of these small-scale motions first grows in time due to the energy transfer from the global mode. Later, they are damped in the presence of resistivity, while they remain undamped in the ideal case \citep[see Figure~6 of][]{terradas2006}.  Thus, normal modes satisfactorily provide a  description of the flux tube global oscillation but do not fully capture the small-scale dynamics in the inhomogeneous layer. The degree to which normal mode eigenfunctions remain a good approximation to the actual plasma motions as time increases is a very relevant question. Previous papers by, e.g., \citet{cally1991} and \citet{cally1997} may be an useful guide to establish the link between the modal analysis of oscillations and the actual time-dependent evolution in inhomogeneous plasmas. 

Here, we focused on studying the effect of transverse density variation, while density stratification along the flux tube was neglected. The effect of longitudinal stratification on standing waves was determined by \citet{andries2005} and \citet{dymova2006} in the TB approximation and by \citet{arregui2005} for fully nonuniform tubes. These authors found that the kink mode period and damping time are weakly dependent on longitudinal stratification, while the ratio of the damping time to the period is completely independent of stratification. The effect of longitudinal stratification on propagating waves was studied by \citet{soler2011strat}, who found that longitudinal stratification can either amplify or attenuate the wave, depending on whether the density decreases or increases toward the direction of wave propagation, respectively. Thus, the amplitude variation of propagating kink waves along stratified tubes is determined by the combined effect of resonant absorption and density stratification. Other effects as, e.g., the presence of mass flows  \citep[e.g.,][]{terradas2010flow,soler2011flow} and magnetic twist \citep[e.g.,][]{terradas2012} can also have some influence on the kink wave damping rate.

Finally, the results of this article suggest that our ignorance about the true density profile in the nonuniform layer might be very relevant for seismology of solar flux tubes  as, e.g., coronal loops, using the observed period and damping rate of transverse oscillations in combination with inversion schemes based on the TTTB approximation \citep{goossens2008seis,goossens2012seis}. These schemes are often used beyond the theoretical range of applicability of the TTTB approximation and ignore the influence of the specific density profile.   Unfortunately, present-day observations do not have enough spatial resolution to determine the shape of the transitional layer, and seismological inversion schemes usually adopt an ad hoc density variation. As shown here,  deviations from the TTTB approximation can be large even for relatively thin nonuniform layers depending on the density profile used.  As a consequence, the error done by the inversion schemes and the reliability of the inferred parameters are uncertain. The impact of the transverse density profile on seismically inferred parameters will be explored in the forthcoming continuation of this article.

\acknowledgements{
We thank J.~L. Ballester and I.~Arregui for reading a draft  of this paper and for giving useful comments. The authors acknowledge support from MINECO and FEDER funds through project AYA2011-22846, and from CAIB through the `Grups Competitius' program and FEDER funds. M.G. acknowledges support from KU Leuven via GOA/2009-009. The research of M.G. has partially been funded by the Interuniversity Attraction Poles Programme initiated by the Belgian Science Policy Office (IAP P7/08 CHARM). J.T. acknowledges support from MINECO through a Ram\'on y Cajal grant. }

\bibliographystyle{apj} 
\bibliography{refs}

\begin{appendix}

\section{Expressions of Coefficients}
\label{app}

The expressions of the coefficients $a_k$ and $s_k$ are as follows,
\begin{eqnarray}
a_0 & = & 1, \\
a_1 & = & -\frac{2 \rho_1 - 2 \rho_2 \ra}{3 \rho_1\ra}a_0, \\
a_2 & = & - \frac{9 \rho_1 \ra a_1 + \left( 2 \rho_1 - 2\rho_2 \ra - 4\rho_3 \ra^2 - m^2 \rho_1 \right) a_0}{8\rho_1 \ra^2}, \\
a_3 & = & -\frac{1}{15 \rho_1 \ra^2} \left[ \left( 4 \rho_2 \ra^2 + 20\rho_1 \ra \right) a_2 + \left( -3 \rho_3 \ra^2 + 3\rho_2 \ra + 6\rho_1 - m^2 \rho_1 \right) a_1 \right. \nonumber \\
 &&+ \left. \left( - 6 \rho_4 \ra^2 - 6\rho_3 \ra +  \frac{\omega^2\ra^2}{B^2/\mu} \rho_1^2 - m^2 \rho_2 \right) a_0 \right], \\
a_4 & = & - \frac{1}{24\rho_1\ra^2} \left[ \sum_{j=0}^3 (j+2)(2j-4)\ra^2 \rho_{5-j} a_j  +\sum_{j=0}^3 (j+2)(4j-5)\ra \rho_{4-j} a_j \right. \nonumber \\
&& + \left. \sum_{j=0}^2 \left( (j+2)(2j-1) -m^2\right) \rho_{3-j}  a_j  + \sum_{j=0}^1 \sum_{l = 0}^{1-j} \frac{\omega^2 \ra^2}{B^2/\mu}  \rho_{l+1} \rho_{2-j-l} a_j + 2\frac{\omega^2 \ra}{B^2 / \mu} \rho_{1}^2 a_0 \right], \\
a_k & = & -\frac{1}{k (k+2)\rho_1\ra^2} \left[ \sum_{j=0}^{k-1} (j+2) (2j - k) \ra^2 \rho_{k-j+1} a_j + \sum_{j=0}^{k-1} (j+2) (4j -2 k + 3) \ra \rho_{k-j} a_j \right. \nonumber \\
&& + \left. \sum_{j=0}^{k-2} \left( (j+2)(2j-k+3) - m^2 \right) \rho_{k-j-1} a_j +  \sum_{j=0}^{k-3} \sum_{l = 0}^{k-j-3} \frac{\omega^2 \ra^2}{B^2/ \mu} \rho_{l+1}\rho_{k-j-l-2} a_j \right. \nonumber \\
&& + \left. \sum_{j=0}^{k-4} \sum_{l = 0}^{k-j-4} 2\frac{\omega^2 \ra}{B^2/ \mu} \rho_{l+1}\rho_{k-j-l-3} a_j  + \sum_{j=0}^{k-5} \sum_{l = 0}^{k-j-5} \frac{\omega^2}{B^2/ \mu} \rho_{l+1}\rho_{k-j-l-4} a_j  \right], \quad \textrm{for} \quad k \geq 5, \\
s_0 &=& 1, \\
s_1 &=& 0, \\
s_2 &=& 0, \\
s_3 &=& \frac{1}{3\rho_1\ra^2} \left[ \left(  m^2\rho_2 - \frac{\omega^2\ra^2}{B^2/\mu} \rho_1^2 \right) s_0 - \mathcal{C} \left( 4\ra^2 \rho_1 a_1 + (\ra^2\rho_2 + 5\ra\rho_1) a_0 \right) \right] , \\
s_4 &=& -\frac{1}{8\rho_1\ra^2}  \left[ 9\rho_1\ra s_3 + \left( 2 \frac{\omega^2\ra}{B^2/\mu}\rho_1^2 + 2 \frac{\omega^2\ra^2}{B^2/\mu} \rho_1\rho_2 - m^2\rho_3  \right) s_0 \right. \nonumber \\
&& + \left. \mathcal{C} \left( 6\ra^2\rho_1 a_2 + 3\ra^2\rho_2 a_1 + 9\ra\rho_1 a_1 + 3(\ra\rho_2 + \rho_1)a_0 \right) \right], \\
s_k &=& -\frac{1}{k(k-2)\rho_1\ra^2} \left\{ \sum_{j=0}^{k-1} j (2j-k-2)\ra^2 \rho_{k-j+1} s_j + \sum_{j=0}^{k-1} j (4j-2k-1)\ra \rho_{k-j} s_j \right. \nonumber \\
&& \left. + \sum_{j=0}^{k-2} \left[ \left( j (2j-k+1) - m^2\right) \rho_{k-j-1} s_j + \mathcal{C} (3j-k+4)\ra^2 \rho_{k-j-1} a_j \right] \right. \nonumber \\
&& + \left. \sum_{j=0}^{k-3} \left[ \sum_{l=0}^{k-j-3} \frac{\omega^2\ra^2}{B^2/\mu} \rho_{l+1} \rho_{k-j-l-2} s_j + \mathcal{C} (6j-2k+11)\ra \rho_{k-j-2} a_j \right] \right. \nonumber \\
&& \left. + \sum_{j=0}^{k-4} \left[ \sum_{l=0}^{k-j-4} 2\frac{\omega^2\ra}{B^2/\mu} \rho_{l+1} \rho_{k-j-l-3} s_j + \mathcal{C} (3j-k+7) \rho_{k-j-3} a_j \right] \right. \nonumber \\
&&  + \left. \sum_{j=0}^{k-5} \sum_{l=0}^{k-j-5} \frac{\omega^2}{B^2/\mu} \rho_{l+1} \rho_{k-j-l-4} s_j  \right\}, \quad \textrm{for} \quad k \geq 5.
\end{eqnarray}

\end{appendix}

\end{document}